\documentclass[aps,eqsecnum,preprint,floats,epsf,epsfig,nofootinbib]{revtex4}
\usepackage{graphicx}
\usepackage{epsfig}

\begin{document}
\def\be{\begin{eqnarray}}
\def\en{\end{eqnarray}}
\def\non{\nonumber}
\def\la{\langle}
\def\ra{\rangle}
\def\pp{{\prime\prime}}
\def\nc{N_c^{\rm eff}}
\def\vp{\varepsilon}
\def\hep{\hat{\varepsilon}}
\def\a{{\cal A}}
\def\B{{\cal B}}
\def\c{{\cal C}}
\def\d{{\cal D}}
\def\e{{\cal E}}
\def\p{{\cal P}}
\def\t{{\cal T}}
\def\up{\uparrow}
\def\dw{\downarrow}
\def\vma{{_{V-A}}}
\def\vpa{{_{V+A}}}
\def\smp{{_{S-P}}}
\def\spp{{_{S+P}}}
\def\J{{J/\psi}}
\def\ov{\overline}
\def\Lqcd{{\Lambda_{\rm QCD}}}
\def\pr{{Phys. Rev.}~}
\def\prl{{ Phys. Rev. Lett.}~}
\def\pl{{ Phys. Lett.}~}
\def\np{{ Nucl. Phys.}~}
\def\zp{{ Z. Phys.}~}
\def\lsim{ {\ \lower-1.2pt\vbox{\hbox{\rlap{$<$}\lower5pt\vbox{\hbox{$\sim$}
}}}\ } }
\def\gsim{ {\ \lower-1.2pt\vbox{\hbox{\rlap{$>$}\lower5pt\vbox{\hbox{$\sim$}
}}}\ } }

\font\el=cmbx10 scaled \magstep2{\obeylines\hfill February, 2010
\obeylines\hfill CYCU-HEP-09-14}

\vskip 1.5 cm

\centerline{\large\bf $B\to V$, $A$, $T$ Tensor Form Factors in
the Covariant Light-Front Approach:}
 \centerline{\large\bf Implications on Radiative $B$ Decays}
\bigskip
\centerline{\bf Hai-Yang Cheng,$^{1,2}$ and Chun-Khiang Chua$^3$}
\medskip
\centerline{$^1$ Institute of Physics, Academia Sinica}
\centerline{Taipei, Taiwan 115, Republic of China}
\medskip
\medskip
\centerline{$^2$ Physics Department, Brookhaven National
Laboratory} \centerline{Upton, New York 11973}
\medskip
\medskip
\centerline{$^3$ Department of Physics, Chung Yuan Christian University}
\centerline{Chung-Li, Taiwan 320, Republic of China}
\medskip


\bigskip
\bigskip
\centerline{\bf Abstract}
\bigskip
\small

We reanalyze the $B\to M$ tensor form factors in a covariant
light-front quark model, where $M$ represents a vector meson $V$, an axial-vector meson $A$, or a tensor meson $T$.
The treatment of masses and mixing angles in the $K_{1A,1B}$
systems is improved, where $K_{1A}$ and $K_{1B}$
are the $^3P_1$ and $^1P_1$ states of the axial-vector meson $K_1$, respectively. Rates of $B\to M\gamma$ decays are then
calculated using the QCD factorization approach. The updated $B\to
K^*\gamma$, $B\to K_1(1270)\gamma$, $K_1(1400)\gamma$ and
$K_2\gamma$ rates agree with the data. The
$K_1(1270)$--$K_1(1400)$ mixing angle is found to be about
$51^\circ$. The sign of the mixing angle is fixed by the observed
relative strength of $B\to K_1(1270)\gamma$ and $K_1(1400)\gamma$.
The formalism is then applied to $B_s\to M$ tensor form factors.
We find
that the calculated $B_s\to \phi\gamma$ rate is consistent with
experiment, though in the lower end of the data. The branching
fractions of $B_s\to f_1(1420)\gamma$ and $f'_2(1525)\gamma$ are
predicted to be of order  $10^{-5}$ and it will be interesting to
search for these modes. Rates on $B_s\to f_1(1285)\gamma$,
$h_1(1380)\gamma$, $h_1(1170)\gamma$, $f_2(1270)\gamma$ decays are
also predicted.

\eject
\section{Introduction}

In this work we shall investigate the $B\to M$ tensor form factors
and their implications on the exclusive radiative $B_{(s)}\to M\gamma$
decays for $\Delta S=1$ transitions with $M$ denoting a vector meson $V$, an axial-vector meson $A$, or a tensor meson $T$.
These decays receive the dominant contributions from the
short-distance electromagnetic penguin process $b\to s\gamma$.
These modes are of great interest since they are loop-induced
processes and are, hence, sensitive to New Physics contributions.
Recently, both CDF~\cite{CDFbetas} and D0~\cite{D0betas} have
observed 1-2 $\sigma$ deviations from the Standard Model (SM)
prediction for the $B_s$--$\bar B_s$ mixing angle. It will be
useful to the search for New Physics in the $B_{u,d,s}$ systems in
the forthcoming experiments at Fermilab, LHCb and Super $B$
factories.

The radiative decay $B\to K^*\gamma$ was first measured by CLEO
\cite{CLEO93} and subsequently updated by CLEO, BaBar and Belle with the results
 \be
 \B(B^0\to K^{*0}\gamma) &=& \cases{ (4.55\pm0.70\pm0.34)\times
 10^{-5} & CLEO \cite{CLEO00} \cr (4.47\pm 0.10\pm0.16)\times
 10^{-5} & BaBar \cite{BaBar09} \cr (4.01\pm0.21\pm0.17)\times
 10^{-5} & Belle \cite{Belle04}, } \non \\
 \B(B^+\to K^{*+}\gamma) &=& \cases{ (3.76\pm0.86\pm0.28)\times
 10^{-5} & CLEO \cite{CLEO00} \cr (4.22\pm0.14\pm0.16)\times
 10^{-5} & BaBar \cite{BaBar09} \cr (4.25\pm0.31\pm0.24)\times
 10^{-5} & Belle \cite{Belle04}. }
 \en
The average branching fractions are \cite{HFAG}
 \be \label{eq:K*BRworld}
 \B(B^0\to K^{*0}\gamma) &=& (4.33\pm0.15)\times 10^{-5}, \non \\
  \B(B^+\to K^{*+}\gamma) &=& (4.21\pm0.18)\times 10^{-5}.
  \en
While the decay $B^-\to K_1(1270)^-\gamma$ has been observed by Belle
in 2004, other $B\to K_1\gamma$ decays have not been
seen and only upper limits were reported~\cite{Belle05}:
 \be
  \B(B^-\to K_1^-(1270)\gamma) &=& (4.3\pm0.9\pm0.9)\times
  10^{-5}, \non \\
 \B(B^-\to K_1^-(1400)\gamma) &<& 1.5\times 10^{-5},
 \non\\
 \B(B^0\to K_1^0(1270)\gamma) &<& 5.8\times 10^{-5},
 \non \\
 \B(B^0\to K_1^0(1400)\gamma) &<& 1.2\times 10^{-5}.
 \en
As for the decay $B\to K_2^*(1430)\gamma$, CLEO \cite{CLEO00} has reported
the first evidence with the combined result of neutral and charged $B$ modes
 \be
 \B(B\to K_2^*\gamma)= (1.66^{+0.59}_{-0.53}\pm0.13)\times
 10^{-5}.
 \en
Later, the Belle measurement \cite{BelleK2} yielded
 \be
 \B(B^0\to K_2^{*0}\gamma)=(1.3\pm0.5\pm0.1)\times 10^{-5},
 \en
while BaBar \cite{BaBarK2} obtained
 \be
  \B(B^0\to K_2^{*0}\gamma) &=& (1.22\pm 0.25\pm 0.10)\times
  10^{-5}, \non \\
 \B(B^+\to K_2^{*+}\gamma) &=& (1.45\pm 0.40\pm 0.15)\times 10^{-5}.
 \en
For radiative $B_s$ decays, Belle has reported the first observation of
$B_s\to\phi\gamma$ decay~\cite{Belle:2007ni} with the result
 \be
 \B(B_s\to\phi\gamma)=(5.7^{+1.8}_{-1.5}{}^{+1.2}_{-1.1})\times 10^{-5}.
 \en
This is the only radiative $B_s$ decay that has been observed so far. Its rate
is similar to those in $B_{u,d}\to K^*\gamma$ decays. Given the fact that
$\tau(B_s)<\tau(B_{u,d})$~\cite{PDG} one will naively expect a
slightly smaller rate for $B_s\to\phi\gamma$.

Using the light-cone sum rule (LCSR) result of
$0.38\pm0.06$~\cite{Ball} for the form factor $T_1(0)$ to be
defined below and the $B\to K^*\gamma$ decay amplitude with
nonfactorizable corrections evaluated in the QCD factorization
(QCDF) approach \cite{BBNS}, it was found in
\cite{QCDfacBFS,QCDfacBB} that the next-to-leading-order (NLO)
corrections will enhance the $B\to K^*\gamma$
rate to the extent that its branching fraction disagrees with the
observed one (\ref{eq:K*BRworld}).

In our previous work~\cite{Cheng:2004yj}, various $B\to M$ tensor
form factors were calculated within the framework of the covariant
light-front (CLF) approach~\cite{Jaus99,CCH}.
This formalism preserves the Lorentz covariance in the light-front
framework and has been applied successfully to describe various properties
of pseudoscalar and vector mesons~\cite{Jaus99}. We extended the
analysis of the covariant light-front model to even-parity,
$p$-wave mesons~\cite{CCH}. Recently, the CLF approach
has been further extended to the studies of the quarkonium system, the
$B_c$ system and so on (see, for example~\cite{CLFext}).

We have pointed out in~\cite{CCH} that relativistic effects could
manifest in heavy-to-light transitions at maximum recoil where the
final-state meson can be highly relativistic and hence there is
no reason to expect that the non-relativistic quark model is still
applicable there. Hence, we believe that the CLF approach can provide
useful information on $B\to M$ transitions at maximum recoil,
the kinematic region relevant to $B\to M\gamma$
decays, and may shed new light on the above-mentioned puzzle.

In \cite{Cheng:2004yj} we showed that a form factor $T_1(0)$ substantially
smaller than what expected from LCSR was obtained and a
significantly improved agreement with experiment was achieved with
the rate calculated using the QCDF method.
Since we have studied $p$-wave mesons before in the CLF
approach~\cite{CCH}, the extension to $B\to K_{1,2}$ transitions,
which could be very difficult for lattice QCD calculations, was
performed straightforwardly and rates on $B\to K_{1,2}\gamma$
decays were predicted using the calculated form factors as
inputs~\cite{Cheng:2004yj}.




In the present work, we revise and extend the analysis
of~\cite{Cheng:2004yj}. We improve the the estimation of the
$K_{1A}$ and $K_{1B}$ mixing angle, where $K_{1A}$ and $K_{1B}$
are the $^3P_1$ and $^1P_1$ states of $K_1$, respectively, and are
related to the physical $K_1(1270)$ and $K_1(1400)$ states. As
will be shown later, the analysis is done consistently within the
covariant light-front approach.
After obtaining tensor form factors in the CLF approach, we use
QCDF as the main theoretical framework to calculate
branching fractions of $B\to K^*\gamma$, $K_1\gamma$ and
$K_2\gamma$ decays.
We further extend our study to radiative decays
$B_s\to\phi\gamma$, $f_1(1420)\gamma$, $f_1(1285)\gamma$,
$h_1(1380)\gamma$, $h_1(1170)\gamma$, $f'_2(1525)\gamma$ and
$f_2(1270)\gamma$. The calculated $B_s\to\phi\gamma$ rate is in
agreement with data. Predictions on the decay rates of other modes are made
and can be checked in future experiments.

The paper is organized as follows. The analytic expressions of the
tensor form factors evaluated in the covariant light-front model
are recollected in Sec. II for completeness. The numerical results
for form factors and decay rates together with discussions are
shown in Sec. III. Conclusion is given in Sec. IV. The formulism
and calculation of the tensor form factors in the covariant
light-front model are shown in Appendix A, while input parameters
for radiative $B$ decay amplitudes in the QCDF approach are
collected in Appendix B.

\section{Tensor form factors}

The matrix element for the $B_q\to M\gamma$ transition with
$M=V,A,T$ mesons is given by
 \be
iM=\la \overline M(P^\pp,\vp^\pp)\gamma(q,\vp)|-iH_{\rm
eff}|\overline
 B_q(P^\prime)\ra,
 \en
where
 \be
 H_{\rm eff}&=&-\frac{G_{\rm F}}{\sqrt2} V^*_{ts} V_{tb}\, c_{7}^{\rm eff} Q_{7},
 \non\\
 Q_{7}&=&\frac{e}{8\pi^2} m_b \bar s \sigma_{\mu\nu} (1+\gamma_5)
 b F^{\mu\nu},
 \en
with $P^{\prime(\pp)}$ being the incoming (outgoing) momentum,
$\vp^{(\pp)}$ the polarization vector of $\gamma$ ($M$), $V_{ij}$
the corresponding Cabbibo-Kobayashi-Maskawa (CKM) matrix element
and $c_{7}^{\rm eff}$ the effective Wilson coefficient. By replacing $c_{7}^{\rm eff}$ by the
effective parameter $a_{7}$, to be discussed below in Sec. III,
nonfactorizable corrections to the $B_q\to M\gamma$ amplitude are
included.
In this work we will update the calculation of the $B\to K^*$ and
$B\to K_1, K^*_2$ transition tensor form factors in the covariant
light-front quark model and extend the study to $B_s\to M\gamma$
decays.

Tensor form factors for $B_q\to V,A,T$ transitions are defined by
  \be
\la \overline V(P^\pp,\vp^\pp)|\bar s\sigma_{\mu\nu}q^\nu
(1+\gamma_5) b|\overline B_q(P^\prime)\ra
          &=& i \epsilon_{\mu\nu\lambda\rho} \vp^{\pp\nu *} P^\lambda q^\rho\, T_1(q^2)
\non\\
           &&  + (\vp^{\pp*}_\mu P\cdot q-P_\mu \vp^{\pp*}\cdot q)T_2(q^2)
\non\\     && +\vp^{\pp*}\cdot q\left(q_\mu-P_\mu\frac{q^2}{P\cdot
                 q}\right) T_3(q^2),
\non\\
\la A_{^3P_1,^1P_1}(P^\pp,\vp^\pp)|\bar s\sigma_{\mu\nu}q^\nu
(1+\gamma_5) b|\overline B_q(P^\prime)\ra
          &=& i \epsilon_{\mu\nu\lambda\rho} \vp^{\pp\nu *} P^\lambda q^\rho\, Y_{A1,B1}(q^2)
\non\\
           &&  + (\vp^{\pp*}_\mu P\cdot q-P_\mu \vp^{\pp*}\cdot q)Y_{A2,B2}(q^2)
\non\\
           &&  +\vp^{\pp*}\cdot q \left(q_\mu-P_\mu\frac{q^2}{P\cdot q}\right) Y_{A3,B3}(q^2),
\non\\
\la T(P^\pp,\vp^\pp)|\bar s\sigma_{\mu\nu}q^\nu (1+\gamma_5)
b|\overline B_q(P^\prime)\ra
          &=& -i \epsilon_{\mu\nu\lambda\rho} \vp^{\pp\nu\sigma *}P^\sigma P^\lambda q^\rho\, \frac
          {U_1(q^2)}{m_{B_q}}
\non\\
           &&  - (\vp^{\pp*}_{\mu\sigma} P\cdot q-P_\mu \vp^{\pp*}_{\sigma\rho}
           q^\rho)P^\sigma\frac{U_2(q^2)}{m_{B_q}}
\non\\
           &&  -\vp^{\pp*}_{\sigma\rho} P^\sigma q^\rho \left(q_\mu-P_\mu\frac{q^2}{P\cdot q}\right)
                \frac{U_3(q^2)}{m_{B_q}},
 \label{eq:ffs}
 \en
where $P=P^\prime+P^\pp$, $q=P^\prime-P^\pp$ and the convention
$\epsilon^{0123}=+1$ is adopted.

\begin{figure}[t!]
\centerline{{\epsfxsize3 in \epsffile{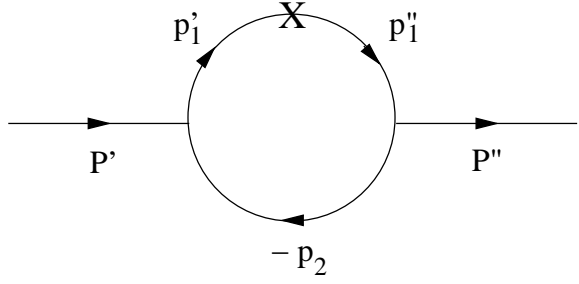}}}
%
\caption{Feynman diagrams for meson transition amplitudes, where
$P^{\prime(\pp)}$ is the incoming (outgoing) meson momentum,
$p^{\prime(\pp)}_1$ is the quark momentum, $p_2$ is the anti-quark
momentum and $X$ denotes the corresponding $\bar q^\pp
\sigma_{\mu\nu} (1+\gamma_5) q^\prime$ transition
vertex.}\label{fig:feyn}
\end{figure}

A brief derivation of $B_q\to V,A,T$ transition tensor form factors
from the diagram depicted in Fig. 1 is shown in Appendix A. Here,
only the final analytic results are given.
First of all, the $B_q\to V$ transition form factors are given
by~\cite{Cheng:2004yj}
 \be
  T_1(q^2)&=&\frac{N_c}{16\pi^3}\int dx_2 d^2 p^\prime_\bot
           \frac{h^\prime_P h^\pp_V}{x_2 \hat N^\prime_1 \hat N^\pp_1}
           \bigg\{2A^{(1)}_1[M^{\prime 2}-M^{\pp2}-2m_1^{\prime2}-2\hat N_1'+q^2
           +2(m_1^\prime m_2+m_1^\pp m_2
  \non\\
           &&-m_1^\prime m_1^\pp)]
           -8 A^{(2)}_1
           +(m_1^{\prime}+m_1^{\pp})^2+\hat N_1'+\hat N_1^\pp-q^2
           +4(M^{\prime 2}-M^{\pp2}) (A^{(2)}_2-A^{(2)}_3)
  \non\\
           &&+4 q^2(-A^{(1)}_1+A^{(1)}_2+A^{(2)}_3-A^{(2)}_4)
           -\frac{4}{w^\pp_V}[(m_1^\prime+m_1^\pp)A^{(2)}_1]\bigg\},
 \non\\
 T_2(q^2)&=&T_1(q^2)+\frac{q^2}{(M^{\prime 2}-M^{\pp2})} \frac{N_c}{16\pi^3}\int dx_2 d^2 p^\prime_\bot
           \frac{h^\prime_P h^\pp_V}{x_2 \hat N^\prime_1 \hat N^\pp_1}
           \bigg\{2 A^{(1)}_2 [M^{\prime 2}-M^{\pp2}-2m_1^{\prime 2}-2\hat N_1'+q^2
 \non\\
           &&+2(m_1^\prime m_2+m_1^\pp m_2-m_1^\prime m_1^\pp)]
           -8 A^{(2)}_1-2M^{\prime 2}+2m_1^{\prime 2}+(m_1^{\prime}+m_1^{\pp})^2
 \non\\
           &&+2(m_2-2m_1')m_2+3\hat N_1'+\hat N_1^\pp-q^2+2 Z_2
           +4 (q^2-2M^{\prime 2}-2 M^{\pp 2}) (A^{(2)}_2-A^{(2)}_3)
 \non\\
           &&-4(M^{\prime 2}-M^{\pp2}) (-A^{(1)}_1+A^{(1)}_2+A^{(2)}_3-A^{(2)}_4)
           -\frac{4}{w^\pp_V}[(m_1^\pp-m_1^\prime+2 m_2)A^{(2)}_1]\bigg\},
 \non\\
 T_3(q^2)&=&\frac{N_c}{16\pi^3}\int dx_2 d^2 p^\prime_\bot
           \frac{h^\prime_P h^\pp_V}{x_2 \hat N^\prime_1 \hat N^\pp_1}
           \bigg\{-2 A^{(1)}_2 [M^{\prime 2}-M^{\pp2}-2m_1^{\prime 2}-2\hat N_1'+q^2
 \\
           &&+2(m_1^\prime m_2+m_1^\pp m_2-m_1^\prime m_1^\pp)]
           +8 A^{(2)}_1+2M^{\prime 2}-2m_1^{\prime 2}-(m_1^{\prime}+m_1^{\pp})^2
 \non\\
           &&-2(m_2-2m_1')m_2-3\hat N_1'-\hat N_1^\pp+q^2
           -2 Z_2
           -4(q^2-M^{\prime 2}-3 M^{\pp 2}) (A^{(2)}_2-A^{(2)}_3)
 \non\\
           &&+\frac{4}{w^\pp_V}\Big\{(m_1^\pp-m_1^\prime+2 m_2)
           [A^{(2)}_1+(M^{\prime 2}-M^{\pp2}) (A^{(2)}_2+A^{(2)}_3-A^{(1)}_1)]
 \non\\
           &&+ (m_1^\prime+m_1^\pp)(M^{\prime 2}-M^{\pp2})
           (A^{(1)}_2-A^{(2)}_3-A^{(2)}_4)
           +m_1^\prime(M^{\prime 2}-M^{\pp2})
           (A^{(1)}_1+A^{(1)}_2-1)
           \Big\}\bigg\}. \non
 \label{eq:T}
 \en
The expressions of $h'$, $h''$, $\hat N'$, $\hat N''$, $A^{(i)}_j$
and $Z_2$ can be found in the Appendix.

Secondly,
the $B_q\to A$ transition form factors can be obtained from the
above expressions by applying a simple relation~\cite{Cheng:2004yj}
(see also Appendix A):
 \be\label{eq:TA}
 Y_{Ai,Bi}(q^2)&=&T_i(q^2) \,\,\,{\rm with}\,\,\,
                         (m_1^\pp\to -m_1^\pp,\,h^\pp_V\to h^\pp_{^3\!A,^1\!A},\,w^\pp_V\to w^\pp_{^3\!A,^1\!A}),
 \en
for $i=1,2,3$. Note that only the $1/w^\pp$ terms in $Y_{Bi}$ form
factors are kept and it should be cautious that the replacement of
$m_1^\pp\to -m_1^\pp$ should not be applied to $m_1^\pp$ in
$w^\pp$ and $h^\pp$.

Thirdly, the $B_q\to T$ transition form factors are given
by~\cite{Cheng:2004yj}
 \be
  U_1(q^2)&=&\frac{N_c}{16\pi^3}\int dx_2 d^2 p^\prime_\bot
           \frac{M' h^\prime_P h^\pp_T}{x_2 \hat N^\prime_1 \hat N^\pp_1}
           \bigg\{2(A^{(1)}_1-A^{(2)}_2-A^{(2)}_3)
           [M^{\prime 2}-M^{\pp2}-2m_1^{\prime2}-2\hat N_1'+q^2
  \non\\
           &&+2(m_1^\prime m_2+m_1^\pp m_2-m_1^\prime m_1^\pp)]
           -8 (A^{(2)}_1-A^{(3)}_1-A^{(3)}_2)+(1-A^{(1)}_1-A^{(1)}_2)
  \non\\
           &&\times[(m_1^{\prime}+m_1^{\pp})^2+\hat N_1'+\hat
           N_1^\pp-q^2]
           +4(M^{\prime 2}-M^{\pp2}) (A^{(2)}_2-A^{(2)}_3-A^{(3)}_3+A^{(3)}_5)
  \non\\
           &&+4 q^2(-A^{(1)}_1+A^{(1)}_2+A^{(2)}_2+A^{(2)}_3-2A^{(2)}_4-A^{(3)}_4+A^{(3)}_6)
           -2(A^{(2)}_1+2A^{(3)}_1-2A^{(3)}_2)
  \non\\
           &&-\frac{8}{w^\pp_V}[(m_1^\prime+m_1^\pp)(A^{(2)}_1-A^{(3)}_1-A^{(3)}_2)]\bigg\},
 \non\\
 U_2(q^2)&=&U_1(q^2)+\frac{q^2}{(M^{\prime 2}-M^{\pp2})}
            \frac{N_c}{16\pi^3}\int dx_2 d^2 p^\prime_\bot
            \frac{M' h^\prime_P h^\pp_T}{x_2 \hat N^\prime_1 \hat N^\pp_1}
            \bigg\{2 (A^{(1)}_2-A^{(2)}_3-A^{(2)}_4)
 \non\\
           &&\times [M^{\prime 2}-M^{\pp2}-2m_1^{\prime 2}-2\hat N_1'+q^2
           +2(m_1^\prime m_2+m_1^\pp m_2-m_1^\prime m_1^\pp)]
 \non\\
           &&
           -8 (A^{(2)}_1-A^{(3)}_1-A^{(3)}_2)
           +(1-A^{(1)}_1-A^{(1)}_2)[-2M^{\prime 2}+2m_1^{\prime 2}+(m_1^{\prime}+m_1^{\pp})^2
 \non\\
           &&+2(m_2-2m_1')m_2+3\hat N_1'+\hat N_1^\pp-q^2]
           +2 [Z_2(1-A^{(1)}_2)-\frac{P\cdot q}{q^2} A^{(2)}_1]
 \non\\
           &&+4 (q^2-2M^{\prime 2}-2 M^{\pp 2}) (A^{(2)}_2-A^{(2)}_3-A^{(3)}_3+A^{(3)}_5)
 \non\\
           &&-4(M^{\prime 2}-M^{\pp2}) (-A^{(1)}_1+A^{(1)}_2+A^{(2)}_2+A^{(2)}_3-2A^{(2)}_4-A^{(3)}_4+A^{(3)}_6)
 \non\\
           &&+2(A^{(2)}_1+2A^{(3)}_1-2A^{(3)}_2)
           -\frac{8}{w^\pp_V}[(m_1^\pp-m_1^\prime+2 m_2)(A^{(2)}_1-A^{(3)}_1-A^{(3)}_2)]\bigg\},
 \non\\
 U_3(q^2)&=&\frac{N_c}{16\pi^3}\int dx_2 d^2 p^\prime_\bot
           \frac{M' h^\prime_P h^\pp_T}{x_2 \hat N^\prime_1 \hat N^\pp_1}
           \bigg\{-2 (A^{(1)}_2-A^{(2)}_3-A^{(2)}_4)
           [M^{\prime 2}-M^{\pp2}-2m_1^{\prime 2}-2\hat N_1'+q^2
 \non\\
           &&+2(m_1^\prime m_2+m_1^\pp m_2-m_1^\prime m_1^\pp)]
           +8 (A^{(2)}_1-A^{(3)}_1-A^{(3)}_2)
           -(1-A^{(1)}_1-A^{(1)}_2)[-2M^{\prime 2}+2m_1^{\prime 2}
 \non\\
           &&+(m_1^{\prime}+m_1^{\pp})^2+2(m_2-2m_1')m_2+3\hat N_1'+\hat N_1^\pp-q^2]
           -2 [Z_2(1-A^{(1)}_2)-\frac{P\cdot q}{q^2} A^{(2)}_1]
 \non\\
           &&-4(q^2-M^{\prime 2}-3 M^{\pp 2}) (A^{(2)}_2-A^{(2)}_3-A^{(3)}_3+A^{(3)}_5)
             -2(A^{(2)}_1+2A^{(3)}_1-2A^{(3)}_2)
 \non\\
           &&+\frac{4}{w^\pp_V}
           \Big\{(m_1^\pp-m_1^\prime+2 m_2)
           [2(A^{(2)}_1-A^{(3)}_1-A^{(3)}_2)+(M^{\prime 2}-M^{\pp2})
 \non\\
           &&\times
           (-A^{(1)}_1+2A^{(2)}_2+2A^{(2)}_3-A^{(3)}_3-2A^{(3)}_4-A^{(3)}_5)]
 \non\\
           &&+ (m_1^\prime+m_1^\pp)(M^{\prime 2}-M^{\pp2})
           (A^{(1)}_2-2A^{(2)}_3-2A^{(2)}_4+A^{(3)}_4+2A^{(3)}_5+A^{(3)}_6)
 \non\\
           &&+m_1^\prime(M^{\prime 2}-M^{\pp2})
           (-1+2A^{(1)}_1+2A^{(1)}_2-A^{(2)}_2-2A^{(2)}_3-A^{(2)}_4)
           \Big\}\bigg\}.
 \label{eq:U}
 \en

We are now ready to calculate the radiative decay rates. Before
proceeding, several remarks are in order: (i) At $q^2=0$, the form
factors obey the simple relations: $T_2(0)=T_1(0)$,
$Y_{A2,B2}(0)=Y_{A1,B1}(0)$ and $U_2(0)=U_1(0)$. (ii)~Form factors
$T_3(0),\,Y_{A3,B3}(0),\,U_3(0)$ do not contribute to the $B\to M\gamma$ radiative decay rates. (iii) There are some new
terms in the above form factor expressions that were missed
in~\cite{Cheng:2004yj}. As we shall see in the next section, the
resulting $B\to M\gamma$ rates are modified sizably for some
modes. It is straightforward to
obtain~\cite{Cheng:2004yj}~\footnote{Since $|V_{cb}V_{cs}^*|\gg
|V_{ub}V_{us}^*|$, for the purpose of obtaining the radiative
decay rates, we only consider the $|V_{cb}V_{cs}^* a_7^c|^2$
contributions.}
 \be
 {\mathcal B}(B_q\to V\gamma)&=&\tau_{B_q}\frac{G_{\rm F}^2 \alpha m_{B_q}^3
                                m_b^2}{32\pi^4}\left(1-\frac{m^2_{V}}{m_{B_q}^2}\right)^3|V_{cb}
                                 V^*_{cs} a^c_{7}\, T_1(0)|^2,
 \non\\
 {\mathcal B}(B_q\to A_{^3P_1,^1P_1}\,\gamma)&=&\tau_{B_q}\frac{G_{\rm F}^2 \alpha m_{B_q}^3
                                m_b^2}{32\pi^4}\left(1-\frac{m^2_{A_{^3P_1,^1P_1}}}{m_{B_q}^2}\right)^3|V_{cb}
                                 V^*_{cs} a^c_{7}\, Y_{A1,B1}(0)|^2,
 \non\\
 {\mathcal B}(B_q\to T\gamma)&=&\tau_{B_q}\frac{G_{\rm F}^2 \alpha m_{B_q}^5
                                m_b^2}{256\pi^4 m_{T}^2}\left(1-\frac{m^2_{T}}{m_{B_q}^2}\right)^5|V_{cb}
                                 V^*_{cs} a^c_{7}\, U_1(0)|^2,
 \label{eq:Br}
 \en
where $\tau_{B_q}$ is the lifetime of the $B_q$ meson and $m_b$ is
the $\overline {\it MS}$ $b$-quark mass. The effective Wilson
coefficient $a_7(V\gamma)$~\cite{QCDfacBFS,QCDfacBB,Ali} and
$a_7(A\gamma)$~\cite{Hatanaka} are calculated in the QCDF
approach~\cite{BBNS}. They consist of several different
contributions \cite{QCDfacBFS,QCDfacBB,Ali,Hatanaka}:
 \be
 a^c_7(\mu)=c^{\rm eff}_7(\mu)+a^c_{7,ver}(\mu)+a^c_{7,sp}(\mu_h)\,,
 \label{eq:a7QCDF1}
 \en
where $c^{\rm eff}_7$, $a_{7,ver}$ and $a_{7,sp}$ are the NLO
Wilson coefficient, the vertex and hard-spectator corrections,
respectively. The last two terms in the above equation are given by
 \be
 a^c_{7,ver}(\mu)&=&\frac{\alpha_s(\mu) C_F}{4\pi}[c_1(\mu)
 G_1(m^2_c/m^2_b)+c^{\rm eff}_8(\mu) G_8],
 \non\\
 a^c_{7,sp}(\mu_h)&=&\frac{\alpha_s(\mu_h) C_F}{4\pi}
              [c_1(\mu_h) H_1(m^2_c/m_b^2)+c_8^{\rm eff}(\mu_h) H_8]
 \label{eq:a7QCDF2}
 \en
with the hadronic scale $\mu_h\sim\sqrt{\Lambda_h\mu}$ for
$\Lambda_h\simeq0.5$~GeV and $G_{1,8}$, $H_{1,8}$ given
in~\cite{QCDfacBB}. Note that the analytic expression for
$a_7(V\gamma)$ and $a_7(A\gamma)$ are identical, but numerically,
due to differences of the wave functions of $V$ and $A$,
$a_{sp}(V\gamma)$ and $a_{sp}(A\gamma)$ could be quite
different~\cite{Hatanaka}. As the QCDF calculation of
$a_7(T\gamma)$ is not available yet, we shall take
 \be
 a^c_7(T\gamma)\simeq c_7^{\rm eff}(\mu)
 \en
and neglect $a^c_{7,ver}(T\gamma)$ and $a^c_{7,sp}(T\gamma)$ in
this work.

In the next section, we will give numerical results for form
factors $T_i(q^2),\,Y_{Ai,Bi}(q^2),\,U_i(q^2)$, and the
corresponding $B_q\to V\gamma,\,A\gamma,\,T\gamma$ decay rates.

\section{Numerical Results and Discussion}

\subsection{$B\to M$ tensor form factors and
$B\to K^*\gamma$, $K_1\gamma$ and $K^*_2\gamma$ decays}

To perform numerical calculations, first we need to  specific some
input parameters in the covariant light-front model. The input
parameters $m_q$ and $\beta$ in the Gaussian-type wave function
(\ref{eq:wavefn}) are shown in Table~\ref{tab:input}. The
constituent quark masses are close to those used in the
literature~\cite{Jaus91,Jaus99,Jaus03,Cheng97,CCH,Hwang02,Cheng:2004yj}.
Meson masses and decay widths are taken from~\cite{PDG} and CKM
parameters from~\cite{CKMfitter}.

The physical $K_1$ states $K_1(1270)$ and $K_1(1400)$ are mixed
states of the $K_{1A}$ and $K_{1B}$ states,
 \be \label{eq:K1mixing}
 K_1(1270)=K_{1A} \sin\theta_{K_1}+K_{1B}\cos\theta_{K_1},
 \nonumber\\
 K_1(1400)=K_{1A} \cos\theta_{K_1}-K_{1B}\sin\theta_{K_1}.
 \en
Since they are
not charge conjugation eigenstates, mixing is
not prohibited. Indeed, the mixing is governed
by the mass difference of the
strange and non-strange light quarks. It follows that the masses
of $K_{1A}$ and $K_{1B}$ can be expressed as
 \be
 m^2_{K_{1A}}&=&m^2_{K_1(1400)} \cos^2\theta_{K_1}+m^2_{K_1(1270)}\sin^2\theta_{K_1},
 \non\\
 m^2_{K_{1B}}&=&m^2_{K_1(1400)}
 \sin^2\theta_{K_1}+m^2_{K_1(1270)}\cos^2\theta_{K_1}.
 \label{eq:mK1AB}
 \en
Note that we need to know the mixing angle $\theta_{K_1}$ in order to specify the mass
parameters $m_{K_{1A,1B}}$, which in turn will be needed to obtain the
numerical results for tensor form factors $Y_{A,B}(q^2)$.

\begin{table}[t!]
\caption{\label{tab:input} The input parameters $m_q$ and $\beta$
(in units of GeV) in the Gaussian-type wave function
(\ref{eq:wavefn}). The parameter $\beta$ for $f_1,h_1,f_2$ is defined for their $s\bar s$ component.
}
\begin{ruledtabular}
\begin{tabular}{cccccc}
          $m_u$
          & $m_s$
          & $m_b$
          & $\beta_B$
          & $\beta_{K^*}$
          & $\beta_{K_1,K^*_2}$
          \\
\hline      $0.25$
          & $0.35$
          & $4.45$
          & $0.5671^{+0.0352}_{-0.0354}$
          & $0.2829$
          & $0.3224_{-0.0195}^{+0.0163}$
          \\
\hline
          $\beta_{B_s}$
          & $\beta_{\phi}$
          & $\beta_{f_1,h_1,f_2}$
          \\
\hline
            $0.6396\pm0.0566$
          & $0.3051$
          & $0.3446\pm0.0064$
\end{tabular}
\end{ruledtabular}
\end{table}

The input parameters $\beta$'s are fixed by the decay constants
whose analytic expressions in the covariant light-front model are
given in \cite{CCH}. We use $f_B=200\pm 15$~MeV,
$f_{B_s}=240\pm15$~MeV, $f_{K^*}=220$~MeV and $f_\phi=230$~MeV to
fix $\beta$'s. For $p$-wave strange mesons, we take for simplicity
$\beta_{K_1}=\beta_{K_{1A}}=\beta_{K_{1B}}=\beta_{K^*_2}$~\cite{ISGW2}.
To fix $\beta_{K_1}$ we need the information of the $K_1(1270)$
and $K_1(1400)$ decay constants.

There exist several estimations on the mixing angle $\theta_{K_1}$
in the literature. From the early experimental information on
masses and the partial rates of $K_1(1270)$ and $K_1(1400)$,
Suzuki found two possible solutions with a two-fold ambiguity,
$|\theta_{K_1}|\approx 33^\circ$ and $57^\circ$~\cite{Suzuki}. A
similar constraint $35^\circ\lsim |\theta_{K_1}|\lsim 55^\circ$
was obtained in \cite{Goldman} based solely on two parameters: the
mass difference of the $a_1$ and $b_1$ mesons and the ratio of the
constituent quark masses. An analysis of $\tau\to
K_1(1270)\nu_\tau$ and $K_1(1400)\nu_\tau$ decays also yielded the
mixing angle to be $\approx 37^\circ$ or $58^\circ$ with a
two-fold ambiguity~\cite{Cheng:DAP}. Most of these estimations
were obtained by assuming a vanishing $f_{K_{1B}}$. With the help
of analytical expressions for $f_{K_{1A,1B}}$ obtained in the  CLF
quark model~\cite{CCH}, we can now release this assumption. Using
the experimental results $\B(\tau\to
K_1(1270)\nu_\tau)=(4.7\pm1.1)\times 10^{-3}$ and $\Gamma(\tau\to
K_1(1270)\nu_\tau)/[\Gamma(\tau\to
K_1(1270)\nu_\tau)+\Gamma(\tau\to
K_1(1400)\nu_\tau)]=0.69\pm0.15$~\cite{PDG}, we obtain
\footnote{The large experimental error with the $K_1(1400)$
production in the $\tau$ decay, namely, $\B(\tau^-\to
K_1^-(1400)\nu_\tau) = (1.7\pm2.6)\times 10^{-3}$ \cite{PDG}, does
not provide sensible information for the $K_1(1400)$ decay
constant.}
 \be
  |f_{K_1(1400)}|=139.2^{+41.3}_{-45.6}~{\rm MeV},
 \quad
 |f_{K_1(1270)}|=169.5^{+18.8}_{-21.2}~{\rm MeV}.
 \label{eq:fK1phys}
 \en
These decay constants are related to $f_{K_{1A}}$ and
$f_{K_{1B}}$ through
 \be \label{eq:K1mixing1}
 m_{K_1(1270)} f_{K_1(1270)}&=&m_{K_{1A}} f_{K_{1A}} \sin\theta_{K_1}
                            +m_{K_{1B}} f_{K_{1B}}\cos\theta_{K_1},
 \nonumber\\
 m_{K_1(1400)} f_{K_1(1400)}&=&m_{K_{1A}} f_{K_{1A}} \cos\theta_{K_1}
                            -m_{K_{1B}} f_{K_{1B}} \sin\theta_{K_1},
 \en
where uses of Eq. (\ref{eq:K1mixing}) and equations for decay
constants $\la
0|A_\mu|K_{1A}\ra=m_{K_{1A}}f_{K_{1A}}\vp_\mu$, $\la
0|A_\mu|K_1(1270)\ra=m_{K_1(1270)}f_{K_1(1270)}\vp_\mu$ and
similar ones for $K_{1B}$ and $K_1(1400)$ have been made. From the
analytic expressions of decay constants given in \cite{CCH}, we see that
$m_{K_{1A}} f_{K_{1A}}$ and $m_{K_{1B}} f_{K_{1B}}$ are functions
of $\beta_{K_1}$ and quark masses only (see Eqs. (2.23) and (2.11)
of \cite{CCH}). In other words, they do not depend on
$m_{K_{1A,1B}}$ and hence $\theta_{K_1}$. Eq. (\ref{eq:K1mixing1})
leads to the relation \be \label{eq:K1mixing2} m_{K_1(1270)}^2
f^2_{K_1(1270)}+m_{K_1(1400)}^2 f^2_{K_1(1400)} =m_{K_{1A}}^2
f_{K_{1A}}^2+ m_{K_{1B}}^2 f_{K_{1B}}^2. \en This relation is
independent of $\theta_{K_1}$. In practice, we shall use this equation
to fix the central value of the parameter $\beta_{K_1}$ to be
$0.3224$ GeV.

Note that in the CLF quark model the signs of the decay constants
$f_{K_{1A}}$ and $f_{K_{1B}}$ and their relative signs with
respect to form factors are fixed~\cite{CCH}.\footnote{The
relative signs of the decay constants, form factors and mixing
angles of the axial-vector mesons were often very confusing in the
literature. As stressed in \cite{CK0708}, the sign of the mixing
angle $\theta_{K_1}$ is intimately related to the relative sign of
the $K_{1A}$ and $K_{1B}$ states. In the light-front quark model
\cite{CCH} and in pQCD \cite{Lu:BtoP}, the decay constants of
$K_{1A}$ and $K_{1B}$ are of opposite signs, while the $D(B)\to
K_{1A}$ and $D(B)\to K_{1B}$ form factors are of the same sign.
The mixing angle $\theta_{K_1}$ is positive. It is the other way
around in the approaches of QCD sum rules \cite{Yang:BtoP} and the
ISGW model \cite{ISGW2}: the decay constants of $K_{1A}$ and
$K_{1B}$ have the same sign, while the $D(B)\to K_{1A}$ and
$D(B)\to K_{1B}$ form factors are opposite in sign. These two
conventions are related via a redefinition of the $K_{1A}$ or
$K_{1B}$ state, i.e., $K_{1A}\to -K_{1A}$ or $K_{1B}\to -K_{1B}$.}
Specifically, we learn from Eq. (2.23) of \cite{CCH}  that
$f_{K_{1A}}$ is negative, whereas $f_{K_{1B}}$ is positive. With
this sign convention, we are ready to determine the mixing angle
$\theta_{K_1}$  from Eq. (\ref{eq:K1mixing1}). We find two best
fit solutions for $\theta_{K_1}$:
 \be
 \theta_{K_1}=\left\{
\begin{array}{ll} 50.8^\circ & {\rm solution~I,} \\
 -44.8^\circ & {\rm solution~II.}  \end{array}\right.
 \en
In both cases,
 \be
 m_{K_{1A}} f_{K_{1A}}=-0.2905~{\rm GeV}^2,
 \quad
 m_{K_{1B}} f_{K_{1B}}= 0.0152~{\rm GeV}^2
 \label{eq:fK1AB}
 \en
are obtained. The uncertainty in $\beta_{K_1}$ for these two mixing
angles can be obtained using Eqs.~(\ref{eq:fK1phys}) and
(\ref{eq:K1mixing1}). The reader may wonder why we do not have a
two-fold ambiguity for $\theta_{K_1}$. This is because we do not
assume a vanishing $f_{K_{1B}}$ and we demand that
$|\theta_{K_1}|\leq\pi/2$. From Eq. (\ref{eq:K1mixing1}) we have
 \be
 \theta_{K_1}&=&\pm \tan^{-1}\left|{ m_{K_1(1270)}f_{K_1(1270)}\over m_{K_1(1400)}f_{K_1(1400)}}\right|
 +\tan^{-1}\left|{m_{K_{1B}}f_{K_{1B}}\over m_{K_{1A}}f_{K_{1A}}}\right|
 \non \\
 &=& \pm47.8^\circ+3.0^\circ.
 \en
This leads to the above two solutions. Note that in the SU(3)
limit, $f_{K_{1B}}=0$ and
$f_{K_1(1270)}/f_{K_1(1400)}=\tan\theta_{K_1}$. As we shall see
below, the second solution $\theta_{K_1}=-44.8^\circ$ is ruled out
by the experimental measurements of $B\to K_1(1270)\gamma$ and
$B\to K_1(1400)\gamma$. For $\theta_{K_1}=50.8^\circ$, we find \be
m_{K_{1A}}=1.37\,{\rm GeV}, && f_{K_{1A}}=-212\,{\rm MeV}, \non \\
m_{K_{1B}}=1.31\,{\rm GeV}, && f_{K_{1B}}=12\,{\rm MeV}. \en

\begin{table}[t]
\caption{Tensor form factors of $B\to K^*,\,K_1,\,K^*_2$
transitions obtained in the covariant light-front model are fitted
to the 3-parameter form Eq. (\ref{eq:FFpara}) except for $Y_{B3}$
and $U_{2,3}$ . Central values of $\beta$'s listed in
Table~\ref{tab:input} are used. All form factors are
dimensionless. For $B\to K_{1A,1B}$ transition form factors, only
results with $\theta_{K_1}=50.8^\circ~$ are shown since one needs
to specify the value of $\theta_{K_1}$ in order to fix the values
of $m_{K_{1A,1B}}$.}
 \label{tab:FF}
\begin{ruledtabular}
\begin{tabular}{| c c c c c || c c c c c |}
~~~$F$~~~~~
    & $F(0)$~~~~~
    & $F(q^2_{\rm max})$~~~~
    & $a$~~~~~
    & $b$~~~~~~
& ~~~ $F$~~~~~
    & $F(0)$~~~~~
    & $F(q^2_{\rm max})$~~~~~
    & $a$~~~~~
    & $b$~~~~~~
 \\
    \hline
$T_1$
    & $0.29$
    & $1.09$
    & 1.86
    & 1.16
&$Y_{A1}$
    & 0.36
    & 1.20
    & 1.61
    & 0.64 \\
$T_2$
    & $0.29$
    & $0.91$
    & 1.03
    & 0.06
&$Y_{A2}$
    & $0.36$
    & $0.58$
    & 0.63
    & $-0.11$
    \\
$T_3$
    & 0.18
    & 0.54
    & 1.48
    & 0.74
&$Y_{A3}$
    & 0.21
    & 0.30
    & 0.76
    & 0.36
    \\
$Y_{B1}$
    & 0.13
    & 0.35
    & 1.88
    & 1.39
&$U_1$
    & 0.28
    & 0.62
    & 2.27
    & 2.33
    \\
 $Y_{B2}$
    & 0.14
    & 0.26
    & 1.00
    & 0.23
&$U_2\,\footnotemark[1]$
    & $0.28$
    & $1.04$
    & $-$
    & $-$
    \\
 $Y_{B3}\,\footnotemark[2]$
    & $-0.05$
    & $-0.17$
    & $2.65$
    & $0.00$
 &$U^{\prime}_2\,\footnotemark[2]$
    & $0.41$
    & $0.78$
    & $1.87$
    & $1.82$
    \\
    &
    &
    &
    &
    &$U_3\,\footnotemark[2] $
    & $-0.25$
    & $-0.68$
    & $-2.27$
    & $1.77$
    \\
\end{tabular}
\end{ruledtabular}
 \footnotetext[1] {We use $U_2\equiv U_1+(q^2/m_B^2) U'_2$ and fit for $U_2'$ using
Eq.~(\ref{eq:FFpara}).}
 \footnotetext[2] {$Y_{B3}$ and $U_3$ are fitted using Eq.~(\ref{eq:FFpara1}).}
\end{table}

Since we have imposed the constraint $q^+=0$ in the calculation, form factors are
obtained only for spacelike momentum transfer $q^2=-q^2_\bot\leq
0$, whereas only the timelike form factors are relevant for the
physical decay processes. Here we
follow~\cite{Jaus91,CCH,Cheng:2004yj} to take the form factors as
explicit functions of $q^2$ in the spacelike region and then
analytically continue them to the timelike region. We find that,
except for the form factors $Y_{B3}$ and $U_{2,3}$, the momentum
dependence of the form factors $T_i,\,Y_{Ai,Bi},\,U_i$ in the
spacelike region can be well parameterized and reproduced in the
three-parameter form:
 \be \label{eq:FFpara}
 F(q^2)=\,{F(0)\over 1-a(q^2/m_{B}^2)+b(q^2/m_{B}^2)^2}\,.
 \en
We then employ this parametrization to determine the physical form
factors at $q^2\geq 0$. In practice, the parameters $a,b$ and
$F(0)$ are obtained by performing a 3-parameter fit to the form
factors in the range $-20\,{\rm GeV}^2\leq q^2\leq 0$. The
obtained $a$ and $b$ coefficients are in most cases not far from
unity as expected. However, the coefficient $b$ for $Y_{B3}$ and
$U_{2,3}$ is rather sensitive to the chosen range for $q^2$ and
can be far away from unity.
To overcome this difficulty, we fit $Y_{B3}(q^2)$ and $U_3(q^2)$
to the form
 \be
\label{eq:FFpara1}
 F(q^2)=F(0)(1+a(q^2/m_{B}^2)+b(q^2/m_{B}^2)^2),
 \en
while for $U_2(q^2)$, we first define $U'_2(q^2)$ through
 \be \label{eq:FFpara2}
 U_2(q^2)&=&U_1(q^2)+\frac{q^2}{m_B^2} U'_2(q^2),
 \en
and then fit $U'_2(q^2)$ using Eq.~(\ref{eq:FFpara}). Note that a
decomposition of $U_2$ into $U_1$ and $U'_2$ is motivated by
Eq.~(\ref{eq:U}). The above procedure accomplishes substantial
improvements.

\begin{figure}[t]
\centerline{
            {\epsfxsize3 in \epsffile{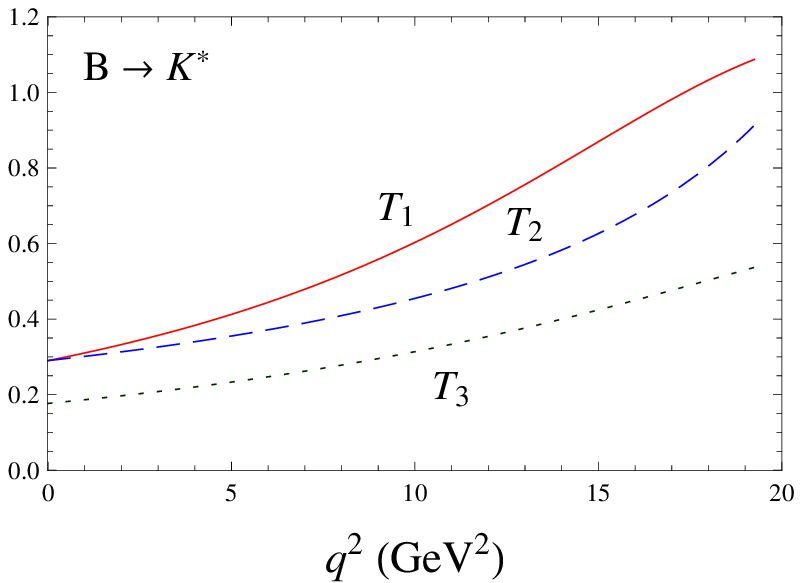}}
            {\epsfxsize3 in \epsffile{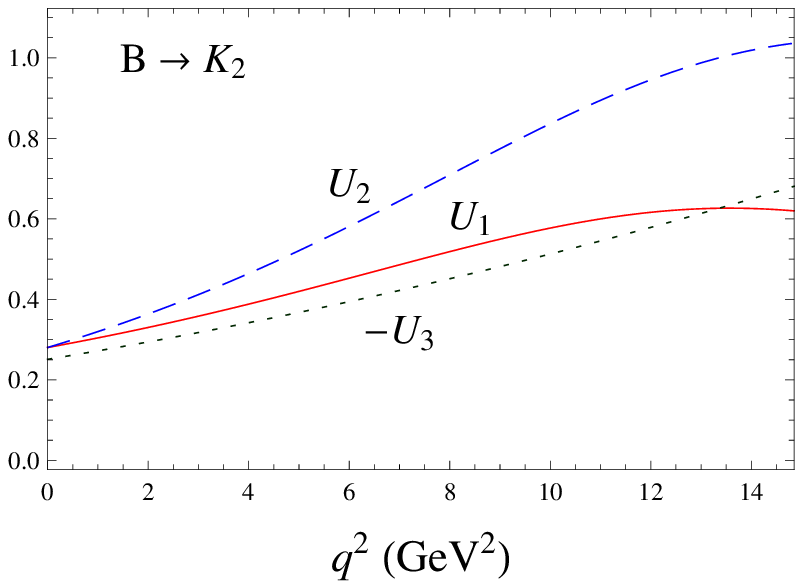}}}
\centerline{
            {\epsfxsize3 in \epsffile{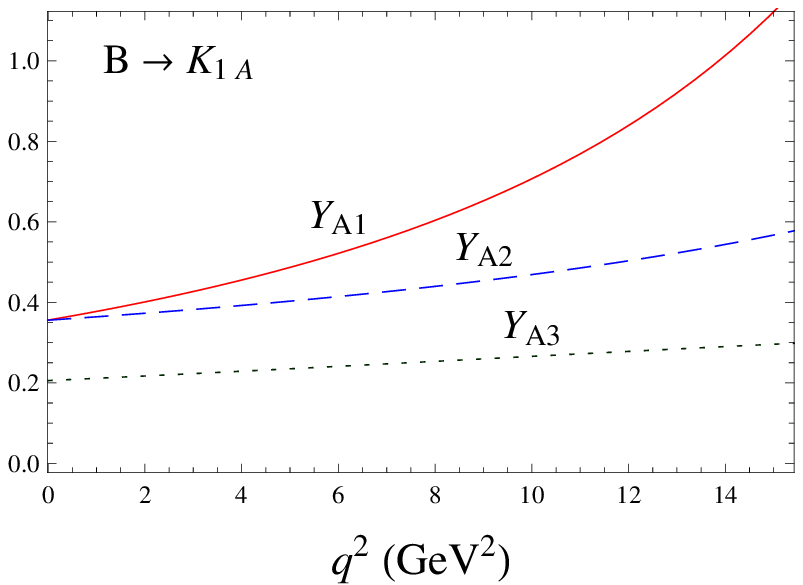}}
            {\epsfxsize3 in \epsffile{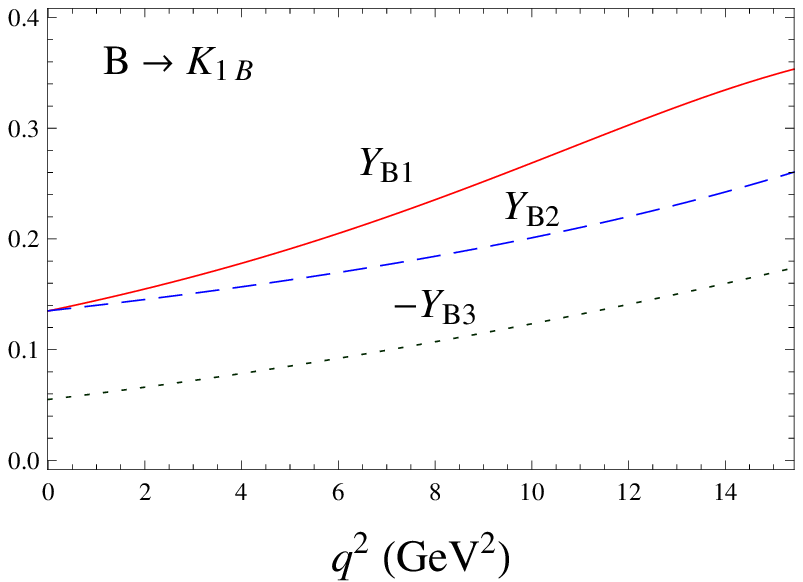}}}
\caption{Tensor form factors $T_{i}(q^2)$, $Y_{Ai,Bi}(q^2)$ and
$U_i(q^2)$ for $B\to K^*$, $B\to K_1$ and $B\to K_2^*$
transitions, respectively.} \label{fig:FF} 
\end{figure}

The tensor form factors and their $q^2$-dependence for $B\to
K^*,\,K_1,\,K^*_2$ transitions are shown in Table~\ref{tab:FF} and
depicted in Fig.~\ref{fig:FF}.
%
%
%
Our form factor $T_1(0)=0.29$ is significantly smaller than the
old light-cone sum rule (LCSR) result of
$0.38\pm0.06$~\cite{Ball}. A new LCSR calculation yields
$0.25^{+0.03}_{-0.02}$ \cite{Hatanaka}, which is close to the
lattice result $T_1(0)=0.24\pm0.03^{+0.04}_{-0.01}$
\cite{Becirevic03}. For the form factors $Y_{A1}$ and $Y_{B1}$ (or
sometimes called $T_1^{K_{1A}}$ and $T_1^{K_{1B}}$, respectively,
in the literature), we compare our results with other model
calculations in Table \ref{tab:YA1}. It is clear that while the
CLF quark model, pQCD \cite{Li} and LCSR \cite{Hatanaka} all lead
to a similar $Y_{A1}$, the predicted $Y_{B1}$ is smaller in
the CLF model.

\begin{table}[t!]
\caption{Tensor form factors $Y_{A1}^{K_{1A}}$ and $Y_{B1}^{K_{1B}}$ at $q^2=0$ in various approaches.} \label{tab:YA1}
\begin{ruledtabular}
\begin{tabular}{c c c c c}
          Form factor
          & This work
          & pQCD \cite{Li}
          & LCSR \cite{Hatanaka}
          & LCSR \cite{Lee}
          \\
\hline
            $Y_{A1}^{K_{1A}}(0)$
          & $0.36\pm0.02$
          & $0.37^{+0.08}_{-0.07}$
          & $0.31^{+0.09}_{-0.05}$
          & --\,\footnotemark[1]
          \\
            $Y_{B1}^{K_{1B}}(0)$
          & $0.13\pm0.01$
          & $0.29^{+0.09}_{-0.09}$
          & $0.25^{+0.06}_{-0.07}$\,\footnotemark[2]
          & $0.256^{+0.0040}_{-0.0044}$
          \\
\end{tabular}
\end{ruledtabular}
\footnotetext[1]{The form factor $Y_{A1}$ was not computed in \cite{Lee}.}
\footnotetext[2]{In our sign convention for $|K_{1}(1270)\ra$ and $|K_{1B}\ra$ states.}
\end{table}

We are now ready to discuss the implications on $B\to M\gamma$ decay
rates. The decay $B\to K^*\gamma$ has been considered in
\cite{QCDfacBFS,QCDfacBB} within the framework of the QCD
factorization approach.  The results of
\cite{QCDfacBFS,QCDfacBB,Ali} are consistent with each other for
the same value of the form factor $T_1(0)$.
For $a_7^c(V\gamma)$ and $a_7^c(A\gamma)$ we shall use Eqs.
(\ref{eq:a7QCDF1}) and (\ref{eq:a7QCDF2}) calculated in QCDF with
input parameters collected in Appendix B. For example, using the formulas given in \cite{QCDfacBB,Hatanaka} and the
central values of input parameters, we obtain
 \be
 a^c_7(m_b)&=&-0.3107+(-0.079-i0.014)
 \non\\
 &&+\frac{f_B f^\bot_M}{m_B
 F^{B\to M}(0)\lambda_B}(\mu_h)[(-0.7906-0.7643i)
 a^\bot_0(\mu_h)
 \non\\
 &&+(-0.2893+0.5024i)a^\bot_1(\mu_h)+(0.1676+0.4252i)a^\bot_2(\mu_h)],
 \label{eq:a7QCDF}
 \en
where contributions from NLO $c^{\rm eff}_7$, $a^c_{7,ver}$ and
$a^c_{7,sp}$ are shown separately and $a_i^\bot$ are Gegenbauer
moments of the meson wave function. The value of $a_7(K^*\gamma)$
is substantially larger than the Wilson coefficient $c_7^{\rm
eff}$ of order $-0.31$ at $\mu=m_b$. For the $K^*_2\gamma$ modes,
we shall employ $a_7=c_7^{\rm eff}$ as NLO QCD corrections from
vertex and hard-spectator contributions there have not been
calculated yet.

\begin{table}[t]
\caption{Branching fractions for the radiative decays $B\to
K^*\gamma$, $K_1(1270)\gamma$, $K_1(1400)\gamma$,
$K^*_2(1430)\gamma$ (in units of $10^{-5}$) in the covariant
light-front model and in other models. Experimental data are taken
from Sec. I.}
 \label{tab:Br}
\begin{ruledtabular}
\begin{tabular}{ l l l l l }
~~~~~~~~~~~~~~~~~~~~
    & $B^-\to K^{*-}\gamma$~~~~~
    & $B^-\to K_1(1270)^-\gamma$~~~~
    & $B^-\to K_1(1400)^-\gamma$~~~~~
    & $B^-\to K^*_2(1430)^-\gamma$~~~~~~
 \\
    \hline
Expt
    & $4.21\pm0.18$
    & $4.3\pm1.2$
    & $<1.5$
    & $1.45\pm0.43$
\\
This work
    & $4.28^{+2.78}_{-1.46}$
    & $5.12^{+1.72}_{-1.77}\,\footnotemark[1]$
    & $0.79^{+0.76}_{-0.25}\,\footnotemark[1]$
    & $2.94^{+3.18}_{-1.39}$
\\
    &
    & $1.26^{+0.99}_{-0.38}\,\footnotemark[2]$
    & $4.50^{+1.53}_{-1.60}\,\footnotemark[2]$
    &
\\
Lattice~\cite{Becirevic03}
    & $2.99^{+2.97}_{-1.13}\,\footnotemark[3]$
    &
    &
    &
    \\
RQM~\cite{EFGT01}
    & $8.2\pm2.7\,\footnotemark[4]$
    & $0.45\pm0.15$
    & $0.78\pm0.18$
    & $1.7\pm0.6$
    \\
LFQM~\cite{GHLZ01}
    & $6.46^{+2.22}_{-1.15}\,\footnotemark[5]$
    &
    &
    &
    \\
LCSR~\cite{Safir01}
    & $3.52\pm1.41\,\footnotemark[6]$
    & $0.71\pm0.28\,\footnotemark[6]$
    & $0.32\pm0.14\,\footnotemark[6]$
    & $1.76\pm0.71\,\footnotemark[6]$
    \\
LCSR~\cite{Hatanaka}
    & $3.22^{+2.38}_{-1.01}\,\footnotemark[7]$
    & $6.6^{+3.7}_{-3.0}\,\footnotemark[8]$
    & $0.65^{+1.28}_{-0.63}\,\footnotemark[8]$
    &
    \\
AP~\cite{Ali}
    & $6.8\pm 2.6$
    &
    &
    &
    \\
BFS~\cite{QCDfacBFS}
    & $7.4^{+0.8}_{-0.9}$\,\footnotemark[9]
    &
    &
    &
    \\
BB~\cite{QCDfacBB}
    & $7.4^{+2.6}_{-2.4}\,\footnotemark[10]$
    &
    &
    &
    \\
BJZ~\cite{BJZ}
    & $5.33\pm1.47$
    &
    &
    &
    \\
HQET~\cite{VO96}
    & $9.99\pm3.81\,\footnotemark[4]$
    & $1.52\pm0.56\,\footnotemark[6]$
    & $0.74\pm0.32\,\footnotemark[6]$
    & $2.18\pm1.02\,\footnotemark[6]$
    \\
 SCET~\cite{SCET}
    & $4.6\pm1.4$
    &
    &
    &
    \\
 PQCD~\cite{Wang:2007an}
    & $3.58^{+1.84}_{-1.35}$
    \\
\end{tabular}
\end{ruledtabular}
 \footnotetext[1]{For the $K_1(1270)$--$K_1(1400)$ mixing angle $\theta_{K_1}=50.8^\circ$.}
 \footnotetext[2]{For the $K_1(1270)$--$K_1(1400)$ mixing angle $\theta_{K_1}=-44.8^\circ$.}
 \footnotetext[3] {Use of $T_1(0)=0.24^{+0.05}_{-0.03}$~\cite{Becirevic03} has been made.}
 \footnotetext[4]{The original result is scaled up by a factor of $|a_7(K^*\gamma)/c_7^{\rm eff}|^2=1.78$\,.}
 \footnotetext[5] {Use of $T_1(0)=0.36$~\cite{GHLZ01} has been made.}
 \footnotetext[6]{Use has been made of ${\mathcal B}(b\to s\gamma)=3.52\times 10^{-4}$~\cite{HFAG}.}
 \footnotetext[7]{Use of $T_1(0)=0.25^{+0.03}_{-0.02}$~\cite{Hatanaka} has been made.}
 \footnotetext[8]{For $\theta_{K_1}=34\pm13^\circ$ in our sign convention for $|K_{1}(1270)\ra$ and $|K_{1B}\ra$ states.}
 \footnotetext[9]{The central value and errors are taken from the complete NLO result for the neutral mode. }
 \footnotetext[10]{For $T_1(0)=0.38$.}
\end{table}

In Table~\ref{tab:Br}, we summarize the calculated branching
fractions for the radiative decays $B\to K^*\gamma$,
$K_1(1270)\gamma$, $K_1(1400)\gamma$, $K^*_2(1430)\gamma$ in the
covariant light-front model. The theoretical errors arise from the
uncertainties in form factors, $a_7$, $|V_{cb} V_{cs}^*|$ and
$m_b$ (see Table~\ref{tab:QCDFinput}). For comparison we also
quote experimental results and some other theoretical
calculations. For results in LFQM~\cite{GHLZ01},
lattice~\cite{Becirevic03} and LCSR~\cite{Safir01}, we also use
Eqs.~(\ref{eq:a7QCDF1}) and (\ref{eq:a7QCDF2}).
For $B\to K^*\gamma$ rates from the relativistic quark model
(RQM)~\cite{EFGT01} and heavy quark effective theory
(HQET)~\cite{VO96}, we have scaled up their results by a factor of
$|a_7(K^*\gamma)/c_7^{\rm eff}|^2=1.78$. Calculations in
LCSR~\cite{Safir01} and HQET~\cite{VO96} are often expressed in
terms of $R\equiv{\mathcal B}(B\to K^{**}\gamma)/{\mathcal B}(b\to
s\gamma)$ with $K^{**}$ denoting $K_1$ or $K_2^*$. Therefore, the
branching fraction of $B\to K^{**}\gamma$ is obtained by multiplying
$R$ with ${\mathcal B}(b\to s\gamma)=3.52\times
10^{-4}$~\cite{HFAG}. Results obtained from large energy effective
theory (LEET)  \cite{Ali}, QCDF  with
long-distance contributions~\cite{BJZ}, soft-collinear effective
theory (SCET)~\cite{SCET} and pQCD~\cite{Wang:2007an} calculations
are also compared. \footnote{The pQCD results for $B\to
K_1(1270)\gamma$ and $K_1(1400)\gamma$ rates in \cite{Wang:2007an}
are not displayed in Table \ref{tab:Br} since the $B\to K_{1A}$
and $B\to K_{1B}$ transition form factors there are erroneous,
though they have been corrected in \cite{Li}.}

As stressed in \cite{QCDfacBFS,QCDfacBB}, the NLO correction
yields an enhancement of the $B\to K^*\gamma$ rate that can be as
large as 80\%. Consequently, the predicted rate will become too
large if the tensor form factor $T_1(0)$ is larger than 0.30. Our
prediction of $\B(B\to K^*\gamma)=(4.28^{+2.78}_{-1.46})\times
10^{-5}$ due to short-distance $b\to s\gamma$ contributions agrees
with experiment (see Table ~\ref{tab:Br}).

From Table ~\ref{tab:Br} we see that our updated $K_1(1270)\gamma$
and $K_1(1400)\gamma$ rates for $\theta_{K_1}=50.8^\circ$ are in
good agreement with the data. Evidently, the other mixing angle
$\theta_{K_1}=-44.8^\circ$  is ruled out by experiment.
As first pointed out in
\cite{Cheng:2004yj}, the $K_1(1400)\gamma$ rate is substantially
smaller than that of $K_1(1270)\gamma$. This can be seen from the
physical form factors \be
Y_1^{K_1(1270)} &=& Y_{A1}\sin\theta_{K_1}+Y_{B1}\cos\theta_{K_1}, \non \\
Y_1^{K_1(1400)} &=& Y_{A1}\cos\theta_{K_1}-Y_{B1}\sin\theta_{K_1}.
\en
It is obvious that the form factor $Y_1$ is large for $K_1(1270)$
and small for $K_1(1400)$ when $\theta_{K_1}=50.8^\circ$.

For $B\to K_2^*\gamma$ decays, the calculated branching fraction
$(2.94^{+3.18}_{-1.39})\times 10^{-5}$ agrees with the world
average of $(1.45\pm0.43)\times 10^{-5}$ within errors.
It should be stressed that the above prediction is for
$a_7(K^*_2\gamma)\simeq c_7^{\rm eff}$. Therefore, a small but
destructive NLO correction will be helpful to improve the
discrepancy.

\subsection{$B_s\to M$ tensor form factors and $B_s\to\phi\gamma$, $h_1\gamma$, $f_1\gamma$ and $f_2\gamma$ decays}

\begin{table}[t!]
\caption{\label{tab:f'f} Summary on mixing angles
 and $m_{s\bar s}$, obtained from Eq.~(\ref{eq:mixing}) and (\ref{eq:mss}),
 for various isosinglet $p$-wave mesons~\cite{CK0708,PDG}.}
\begin{ruledtabular}
\begin{tabular}{cccccc}
          $^{2s+1}l_J$
          & $f'$
          & $f$
          & $\alpha(^\circ)$
          & $m_{s\bar s}$(GeV)
          \\
\hline
            $^1P_1$
          & $h_1(1380)$
          & $h_1(1170)$
          & $54.7$
          & $1.32$
          \\
            $^3P_1$
          & $f_1(1420)$
          & $f_1(1285)$
          & $94.9$
          & $1.43$
          \\
            $^3P_2$
          & $f'_2(1525)$
          & $f_2(1270)$
          & $84.3$
          & $1.52$
\end{tabular}
\end{ruledtabular}
\end{table}

\begin{table}[t!]
\caption{Same as Table~\ref{tab:FF} except for the tensor
form factors of $B_s\to
\phi,\,f^{(\prime)}_1,\,h^{(\prime)}_1,\,f_2$ transitions. Note
that Clebsch-Gordan coefficients are not included (see the text
for more details).}
 \label{tab:BsFF}
\begin{ruledtabular}
\begin{tabular}{| c c c c c || c c c c c |}
~~~$F$~~~~~
    & $F(0)$~~~~~
    & $F(q^2_{\rm max})$~~~~
    & $a$~~~~~
    & $b$~~~~~~
& ~~~ $F$~~~~~
    & $F(0)$~~~~~
    & $F(q^2_{\rm max})$~~~~~
    & $a$~~~~~
    & $b$~~~~~~
 \\
    \hline
$T_1$
    & $0.27$
    & $0.72$
    & 1.99
    & 1.58
&$Y_{A1}$
    & 0.36
    & 1.07
    & 1.70
    & 0.89
    \\
$T_2$
    & $0.27$
    & $0.91$
    & 1.17
    & 0.18
&$Y_{A2}$
    & $0.36$
    & $0.58$
    & 0.67
    & $-0.06$
    \\
$T_3$
    & 0.16
    & 0.40
    & 1.54
    & 0.96
&$Y_{A3}$
    & 0.23
    & 0.35
    & 0.90
    & 0.48
    \\
$Y_{B1}$
    & 0.12
    & 0.29
    & 1.98
    & 1.73
&$U_1$
    & 0.28
    & 0.55
    & 2.30
    & 2.65
    \\
 $Y_{B_2}$
    & 0.12
    & 0.28
    & 1.17
    & 0.37
&$U_2{}^a$
    & $0.28$
    & $0.78$
    & $-$
    & $-$
    \\
 $Y_{B3}{}^b$
    & $-0.09$
    & $-0.24$
    & $2.23$
    & $0.01$
 &$U^{\prime}_2{}^a$
    & $0.29$
    & $0.45$
    & $2.10$
    & $2.75$
    \\
    &
    &
    &
    &
    &$U_3{}^b$
    & $-0.18$
    & $-0.55$
    & $2.74$
    & $0.07$
    \\
\end{tabular}
\end{ruledtabular}
  \footnotetext[1] {We use $U_2\equiv U_1+(q^2/m_{B_s}^2) U'_2$ and fit for $U_2'$ using
Eq.~(\ref{eq:FFpara}).}
 \footnotetext[2] {$Y_{B3}$ and $U_3$ are fitted using Eq.~(\ref{eq:FFpara1}).}
\end{table}

We use $f_{B_s}=240\pm15$~MeV and $f_\phi=230$~MeV to fix the input parameters $\beta_{B_s}$ and $\beta_{\phi}$, respectively. For $p$-wave mesons, there are mixing between
singlet and octet states or, equivalently, between $u\bar u+d\bar
d$ and $s\bar s$ components, where only the $s\bar s$ components
are relevant to $B_s\to M\gamma$ transitions. We follow~\cite{PDG}
to use
 \be
 f'&=&\frac{1}{\sqrt2}(u\bar u+d\bar d)\cos\alpha-s\bar
 s\sin\alpha,
 \non\\
 f&=&\frac{1}{\sqrt2}(u\bar u+d\bar d)\sin\alpha+s\bar
 s\cos\alpha,
 \label{eq:alpha}
 \en
with $(f',f)=(h_1(1380), h_1(1170))$ for $^1P_1$ states,
$(f_1(1420), f_1(1285))$ for $^3P_1$ states and $(f'_2(1525),
f_2(1270))$ for $^3P_2$ tensor states~\cite{PDG}.
The mixing
angle $\alpha$ is related to the singlet-octet
mixing angle $\theta$ by the relation $\alpha=\theta+54.7^\circ$. The latter mixing angle is defined by
\be
 f'&=&f_8\cos\theta-f_1\sin\theta,
 \non\\
 f&=& f_8\sin\theta+f_1\cos\theta,
\en
and determined by the mass relations~\cite{PDG,CK0708}
 \be
 \tan^2\theta_{^3P_1}&=&\frac{4 m_{K_{1A}}^2-m_{a_1}^2-3 m_{f_1(1420)}^2}
                             {-4 m_{K_{1A}}^2+m_{a_1}^2+3 m_{f_1(1285)}^2},
 \non\\
 \tan^2\theta_{^1P_1}&=&\frac{4 m_{K_{1B}}^2-m_{b_1}^2-3 m_{h_1(1380)}^2}
                             {-4 m_{K_{1B}}^2+m_{b_1}^2+3 m_{h_1(1170)}^2},
 \label{eq:mixing}
 \en
derived from the Gell-mann-Okubo mass formula,
where $m_{K_{1A,1B}}$ can be inferred from Eq.~(\ref{eq:mK1AB})
with $\theta_{K_1}=50.8^\circ$. The signs of these angles can be
determined from~\cite{CK0708,PDG}
 \be
 \tan\theta_{^3P_1}&=&\frac{4 m_{K_{1A}}^2-m_{a_1}^2-3 m_{f_1(1420)}^2}
                           {2\sqrt2 (m^2_{a_1}-m^2_{K_{1A}})},
 \non\\
 \tan\theta_{^1P_1}&=&\frac{4 m_{K_{1B}}^2-m_{b_1}^2-3 m_{h_1(1380)}^2}
                           {2\sqrt2 (m^2_{b_1}-m^2_{K_{1B}})}.
 \en
Denoting the mass of the $\bar ss$ component as $m_{s\bar s}$, we have
 \be
 m_{s\bar s}^2=m_{f'}^2 \sin^2\alpha+m_f^2\cos^2\alpha,
 \label{eq:mss}
 \en
The obtained $m_{s\bar s}$ for various states are summarized in
Table~\ref{tab:f'f}.

Defining $\la 0|\bar s\gamma_\mu\gamma_5 s|s\bar s\ra=m_{s\bar s}f^s\vp_\mu$
and $\la 0|\bar s\gamma_\mu\gamma_5 s|f\ra=m_ff^s_f\vp_\mu$, it follows
from Eq.~(\ref{eq:alpha}) that
\be
m_{f'}f^s_{f'}=-m_{s\bar s}f^s\sin\alpha, \qquad m_{f}f^s_{f}=m_{s\bar s}f^s\cos\alpha.
\en
From the values of $\alpha$ and $m_{s\bar s}$ shown in Table~\ref{tab:f'f}
and the decay constants of $f_1(^3P_1)$ and $f_8(^3P_1)$ determined to be $-245\pm13$~MeV and $-239\pm
13$~MeV, respectively, in \cite{Yang:2007zt}, we obtain $f^s(^3P_1)= f^s_{f(1420)}
m_{f(1420)}/(-m_{s\bar s}\sin\alpha)= -230\pm9$~MeV,~{\footnote
{Using $f^s(^3P_1)= f^s_{f(1285)} m_{f(1285)}/(m_{s\bar
s}\cos\alpha)$, a similar central value is obtained, but the error
is of order 100 MeV.}}
which is the decay constant of the $^3P_1$
axial vector meson with a pure $s\bar s$ quark content.
Consequently, $\beta_{f_1,s\bar s}$ is determined and shown in
Table~\ref{tab:input}. For $p$-wave mesons, we take for simplicity
$\beta_{f_1,s\bar s}=\beta_{h_1,s\bar
s}=\beta_{f_2,s\bar s}$~\cite{ISGW2}. Input parameters relevant to
$B_s\to M\gamma$ decays are summarized in Table~\ref{tab:input}.

\begin{figure}[t!]
\centerline{
            {\epsfxsize3 in \epsffile{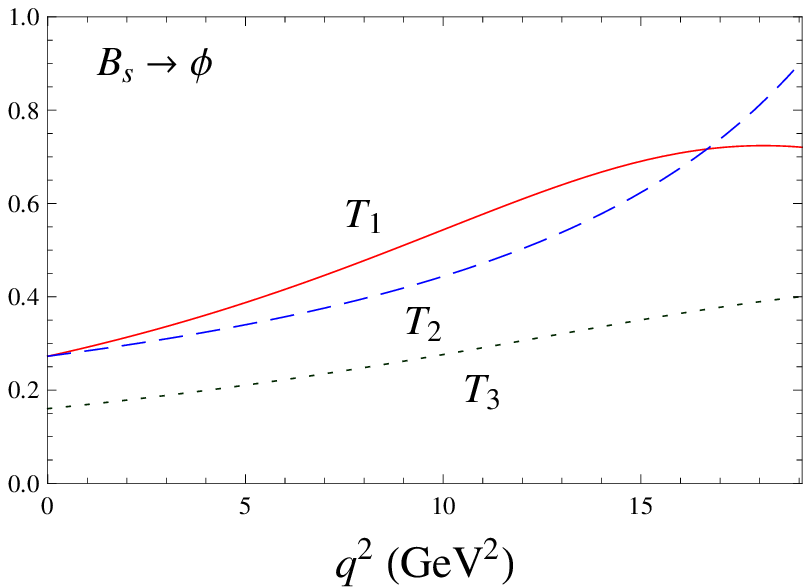}}
            {\epsfxsize3 in \epsffile{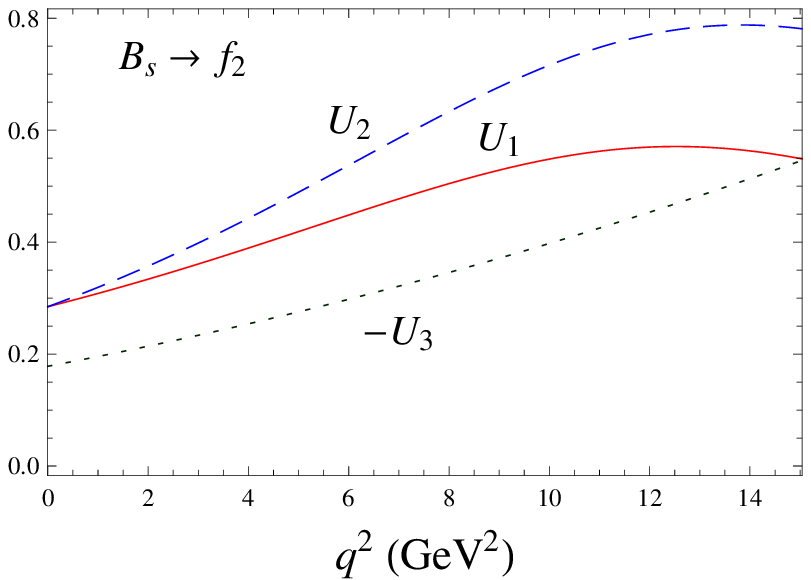}}}
\centerline{
            {\epsfxsize3 in \epsffile{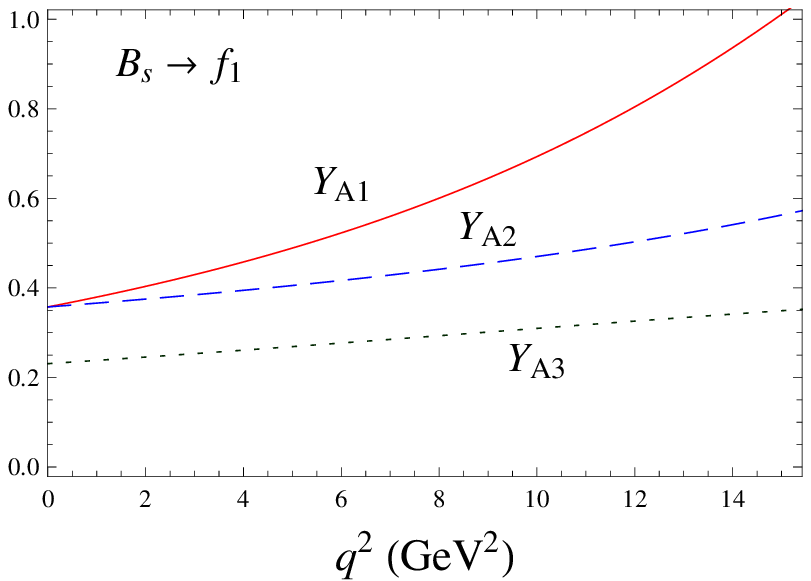}}
            {\epsfxsize3 in \epsffile{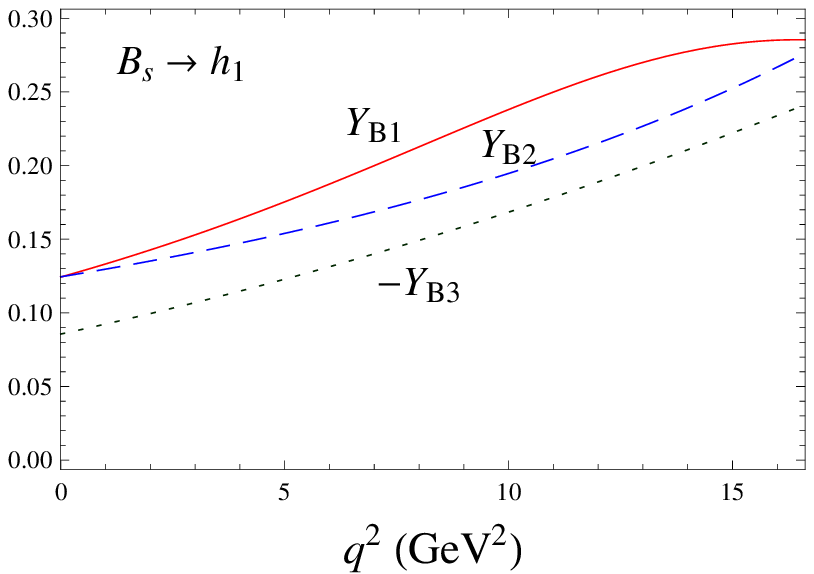}}}
\caption{Same as Fig. 2 except for $B_s\to M$ transitions.}
\label{fig:FFBs} 
\end{figure}

Tensor form factors for $B_s\to V, A(^3P_1), A(^1P_1), T(^3P_2)$
transitions are shown in Table~\ref{tab:BsFF}. As in the $B$ decay
case, except for the form factors $Y_{B3}$ and $U_{2,3}$, the
momentum dependence of the form factors $T_i,\,Y_{Ai,Bi},\,U_i$
are fitted to the three-parameter form given in Eq.
(\ref{eq:FFpara}) with $m_B$ replaced by $m_{B_s}$, while
$Y_{B3}(q^2)$, $U'_2(q^2)$ and $U_3(q^2)$ are fitted to the form
shown in Eq.~(\ref{eq:FFpara1}) with $m_B$ replaced by $m_{B_s}$,
as well. Recall that $U'_2$ is defined through
Eq.(\ref{eq:FFpara2}). These form factors are plotted in
Fig.~\ref{fig:FFBs}. Comparing Tables~\ref{tab:FF} and
\ref{tab:BsFF}, we notice that the values of form factors at $q^2=0$
are similar to the corresponding ones in $B$ transitions.
Therefore, flavor of the spectator quark does not seem to play
a special role in these radiative $B$ and $B_s$ decays.

Form factors for $B_s\to f_1,h_1,f^{(\prime)}_2$ transitions with
physical final states can be obtained from Table~\ref{tab:BsFF} by
including suitable Clebsch-Gordan coefficients. Specifically, form
factors for various $B_s\to M$ transitions with $i=1,2,3$ are
given by
 \be
 Y_{f_1(1420)i}=-\sin\alpha_{^3P_1}\times Y_{Ai},
 &\quad&
 Y_{f_1(1285)i}=\cos\alpha_{^3P_1}\times Y_{Ai},
 \non\\
 Y_{h_1(1380)i}=-\sin\alpha_{^1P_1}\times Y_{Bi},
 &\quad&
 Y_{h_1(1170)i}=\cos\alpha_{^1P_1}\times Y_{Bi},
 \non\\
 Y_{f'_2(1525)i}=-\sin\alpha_{^3P_2}\times U_{i},
 &\quad&
 Y_{f_2(1270)i}=\cos\alpha_{^3P_2}\times U_{i}.
 \en
Since only the $s\bar s$ components of these mesons can be
transited from a $B_s$ meson via a $\bar b\sigma_{\mu\nu} s$
density, the sizes of the corresponding form factors are reduced
by the Clebsch-Gordan coefficients (see also
Eq.~(\ref{eq:alpha})).

\begin{table}[t]
\caption{Branching fractions for the radiative decays $B_s\to
\phi\gamma$, $f_1(1420)\gamma$, $f_1(1285)\gamma$,
$h_1(1380)\gamma$, $h_1(1170)\gamma$, $f'_2(1525)\gamma$,
$f_2(1270)\gamma$ (in units of $10^{-5}$) in the covariant
light-front model and other models. Experimental data are from
\cite{Belle:2007ni,HFAG}.}
 \label{tab:BsBr}
\begin{ruledtabular}
\begin{tabular}{ l l l l l }
~~~~~~~~~~~~~~~~~~~~
    & $B_s\to \phi\gamma$~~~~~
    & $B_s\to f_1(1420)\gamma$~~~~
    & $B_s\to f_1(1285)\gamma$~~~~~
    & $B_s\to h_1(1380)\gamma$
   \\
    \hline
Expt
    & $5.7^{+2.1}_{-1.8}$
    &
    &
    &
\\
This work
    & $3.39^{+2.45}_{-1.22}$
    & $4.81^{+1.55}_{-1.17}$
    & $0.03^{+0.11}_{-0.01}$
    & $0.27^{+0.14}_{-0.15}$
   \\
BJZ~\cite{BJZ}
    & $3.94\pm1.19$
    &
    &
    &
    \\
 SCET~\cite{SCET}
    & $4.3\pm1.4$
    &
    &
    &
    \\
 PQCD~\cite{Wang:2007an}
    & $3.58^{+1.46}_{-1.09}$
    & $6.19^{+3.06}_{-2.52}\,\footnotemark[1]$
    & $0.01^{+0.01}_{-0.01}\,\footnotemark[1]$
    & $4.44^{+2.09}_{-1.66}\,\footnotemark[3]$
    \\
    &
    & $5.82^{+2.88}_{-2.38}\,\footnotemark[2]$
    & $0.38^{+0.18}_{-0.14}\,\footnotemark[2]$
    & $5.00^{+2.22}_{-1.85}\,\footnotemark[4]$
    \\
\hline
    & $B_s\to h_1(1170)\gamma$~~~~
    & $B_s\to f'_2(1525)\gamma$~~~~~
    & $B_s\to f_2(1270)\gamma$
    \\
\hline
This work
    & $0.15^{+0.07}_{-0.08}$
    & $2.30^{+2.19}_{-0.99}$
    & $0.04^{+0.04}_{-0.02}$
    \\
PQCD~\cite{Wang:2007an}
    & $0.79^{+0.36}_{-0.28}\,\footnotemark[3]$
    &
    &
    \\
    & $0.23^{+0.12}_{-0.01}\,\footnotemark[4]$
    &
    &
    &
    \\
\end{tabular}
\end{ruledtabular}
 \footnotetext[1]{For the mixing angle $\theta_{^3P_1}=38^\circ$.}
 \footnotetext[2]{For the mixing angle $\theta_{^3P_1}=50^\circ$.}
 \footnotetext[3]{For the mixing angle $\theta_{^1P_1}=10^\circ$.}
 \footnotetext[4]{For the mixing angle $\theta_{^1P_1}=45^\circ$.}
 \end{table}

For the effective Wilson coefficient $a_7$, we shall use
the QCDF ones as shown in  Eqs. (\ref{eq:a7QCDF1}) and
(\ref{eq:a7QCDF2}) with input parameters given in Appendix B.

Rates of radiative  $B_s\to \phi\gamma$, $f_1(1420)\gamma$,
$f_1(1285)\gamma$, $h_1(1380)\gamma$, $h_1(1170)\gamma$,
$f'_2(1525)\gamma$, $f_2(1270)\gamma$ decays can be obtained in
analog to the $B$ meson case. Results obtained by using tensor
form factors calculated in the covariant light-front model are
shown in Table~\ref{tab:BsBr} where comparison with results from other
models~\cite{BJZ,SCET,Wang:2007an} and data~\cite{Belle:2007ni} is also made.
We see that the calculated $B_s\to\phi\gamma$ rate is consistent
with the data~\cite{Belle:2007ni} and other models
\cite{BJZ,SCET,Wang:2007an} within errors. Note that our
$B_s\to\phi\gamma$ branching fraction is smaller than the $B\to
K^*\gamma$ one. The branching fraction of $B_s\to \phi\gamma$ can
be related to the $B\to K^*\gamma$ one via
 \be
\B(B_s\to \phi\gamma) &=& \left( {m_B\over m_{B_s}}\right)^3
\left( {m_{B_s}^2-m_\phi^2\over m_B^2-m_{K^*}^2} \right)^3\,
{\tau(B_s)\over \tau(B)}
\left| {a^c_7(\phi\gamma)\,T_1^{B_s\phi}(0)\over a^c_7\,(K^*\gamma)T_1^{BK^*}(0)}\right|^2 \B(B\to K^*\gamma)  \non \\
&\simeq& 0.914\left| {T_1^{B_s\phi}(0)\over
T_1^{BK^*}(0)}\right|^2 \B(B\to K^*\gamma).
 \en
It is clear that the reduction arises from
the fact that $T_1(0)$ for the $B_s\to\phi$ transition is smaller
than that for the $B\to K^*$ one by 7\% and the ratio of $B_s$ and
$B$ lifetimes $\tau(B_s)/\tau(B)\simeq 0.87$~\cite{PDG} leads to a
further suppression.

Branching fractions for $B_s\to f_1(1420)\gamma$ and $f'_2(1525)\gamma$ are predicted to reach the level of $10^{-5}$. It will be interesting to
search for these modes in the near future. Comparing to other
predictions, we note that most of our results on $B_s\to A\gamma$
decays agree with those in \cite{Wang:2007an} except the one in
$B_s\to h_1(1380)\gamma$ decay, where our result is about one
order of magnitude smaller. Our predictions on $B_s\to
f_1(1420)\gamma$, $f_1(1285)\gamma$, $h_1(1380)\gamma$,
$h_1(1170)\gamma$, $f'_2(1525)\gamma$, $f_2(1270)\gamma$ rates can
also be checked in future experiments.

\section{Conclusion}

$B\to M$ and $B_s\to M$ tensor form factors are calculated in the
covariant light-front quark model. All numerical results are
analyzed using the CLF formulas in \cite{Cheng:2004yj} with
previously missing terms being included (see the erratum of
\cite{Cheng:2004yj}). Exclusive radiative $B$ and $B_s$ decays,
$B\to K^*\gamma$, $K_1(1270)\gamma$, $K_1(1400)\gamma$,
$K_2^*(1430)\gamma$ and $B_s\to f_1(1420)\gamma$,
$f_1(1285)\gamma$, $h_1(1380)\gamma$, $h_1(1170)\gamma$,
$f'_2(1525)\gamma$, $f_2(1270)\gamma$, are obtained using QCDF.
Our main conclusions are as follows:

 \begin{enumerate}

 \item
The treatment on $m_{K_{1A}}$ and $m_{K_{1B}}$ is improved. In
\cite{Cheng:2004yj} these masses were determined with some
approximations from the measured masses of $K_1(1270)$,
$K_1(1400)$, $b_1(1232)$ and $h_1(1380)$ and no information of the
mixing angle was used. In the present work, we use
Eq.~(\ref{eq:mK1AB}) to determine these masses. This procedure
does not rely on any
approximation. 

\item The treatment on the $K_{1A}-K_{1B}$ mixing angle $\theta_{K_1}$
is also improved. In~\cite{Cheng:2004yj},  $\theta_{K_1}$ was taken to
be $\pm 37$ and $\pm 58$ degrees from other analyses. These
analyses were either based on the assumption of a vanishing decay
constant of $K_{1B}$ or relied on some other calculated results
of $f_{K_{1A}}$. Since the formalism employed in this work is
capable of providing information on $f_{K_{1A}}$ and $f_{K_{1B}}$,
we can analyze the mixing angle consistently within the covariant
light front approach.

\item $B \to V\gamma$ and $A \gamma$ decay rates are obtained
using the QCDF approach with form factors calculated in this work.
The predictions on $B \to A \gamma$ rates are more reliable than
that in~\cite{Cheng:2004yj}, where only a na\"{i}ve estimation on
the effective Wilson coefficients was used.

 \item
The updated $B\to K_1(1270)\gamma$ rate is in agreement with the
data, while the $B\to K_1(1400)\gamma$ rate is consistent with the
experimental bound [8]. These decay rates are very sensitive to
the $K_1(1270)$--$K_1(1400)$ mixing angle and we found that
$\theta_K=50.8^\circ$ is favored by the data.

 \item
 The predicted $B\to K^*\gamma$ and $K_2\gamma$ rates agree with data.

 \item
The calculated $B_s\to \phi\gamma$ rate agree with experiment,
though in the lower end of the data.

\item  In addition, we have studied all $B_s \to (A,T) \gamma$
decays with $b \to s$ transition. Branching fractions of $B_s\to
f_1(1420)\gamma$ and $f'_2(1525)\gamma$ are predicted to reach the
level of $10^{-5}$. It will be interesting to search for these
modes. Our predictions on $f_1(1285)\gamma$, $h_1(1380)\gamma$,
$h_1(1170)\gamma$, $f_2(1270)\gamma$ decay rates can also be
checked in future experiments.

 \end{enumerate}

\vskip 2.5cm \acknowledgments


One of us (H.Y.C.) wishes to thank the hospitality of the Physics
Department, Brookhaven National Laboratory. This research was
supported in part by the National Science Council of R.O.C. under
Grant Nos. NSC97-2112-M-001-004-MY3 and NSC97-2112-M-033-002-MY3.

\newpage

\appendix
\section{A Brief derivation of analytical expressions of tensor form factors}

In this appendix, we give brief derivation that leads to the
analytic formulas of tensor form factors given in
\cite{Cheng:2004yj}. We consider the transition amplitude given by
the one-loop diagram as shown in Fig.~\ref{fig:feyn}.
The incoming (outgoing) meson has the momentum
$P^{\prime(\pp)}=p_1^{\prime(\pp)}+p_2$, where $p_1^{\prime(\pp)}$
and $p_2$ are the momenta of the off-shell quark and antiquark,
respectively, with masses $m_1^{\prime(\pp)}$ and $m_2$. These
momenta can be expressed in terms of the internal variables $(x_i,
p_\bot^\prime)$,
 \be
 p_{1,2}^{\prime+}=x_{1,2} P^{\prime +},\qquad
 p^\prime_{1,2\bot}=x_{1,2} P^\prime_\bot\pm p^\prime_\bot,
 \en
with $x_1+x_2=1$. Note that we use $P^{\prime}=(P^{\prime -},
P^{\prime +}, P^\prime_\bot)$, where $P^{\prime\pm}=P^{\prime0}\pm
P^{\prime3}$, so that $P^{\prime 2}=P^{\prime +}P^{\prime
-}-P^{\prime 2}_\bot$.
In the covariant light-front approach,
total four momentum is conserved at each vertex where quarks and
antiquarks are off-shell. It is useful to define some internal
quantities:
\begin{eqnarray} \label{eq:internalQ}
 M^{\prime2}_0
          &=&(e^\prime_1+e_2)^2=\frac{p^{\prime2}_\bot+m_1^{\prime2}}
                {x_1}+\frac{p^{\prime2}_{\bot}+m_2^2}{x_2},\quad\quad
                \widetilde M^\prime_0=\sqrt{M_0^{\prime2}-(m^\prime_1-m_2)^2},
 \nonumber\\
 e^{(\prime)}_i
          &=&\sqrt{m^{(\prime)2}_i+p^{\prime2}_\bot+p^{\prime2}_z},\quad\qquad
 p^\prime_z=\frac{x_2 M^\prime_0}{2}-\frac{m_2^2+p^{\prime2}_\bot}{2 x_2 M^\prime_0}.
 \end{eqnarray}
Here $M^{\prime2}_0$ can be interpreted as the kinetic invariant
mass squared of the incoming $q\bar q$ system, and $e_i$ the
energy of the quark $i$.

\begin{table}[b]
\caption{\label{tab:feyn} Feynman rules for the vertices
($i\Gamma^\prime_M$) of the incoming mesons-quark-antiquark, where
$p^\prime_1$ and $p_2$ are the quark and antiquark momenta,
respectively. Under the contour integrals to be discussed below,
$H^\prime_M$ and $W^\prime_M$ are reduced to $h^\prime_M$ and
$w^\prime_M$, respectively, whose expressions are given by
Eq.~(\ref{eq:h}). Note that for outgoing mesons, we shall use
$i(\gamma_0\Gamma^{\prime\dagger}_M\gamma_0)$ for the
corresponding vertices.}
\begin{tabular}{|c| c|}
\hline
 $M\,(^{2S+1}L_J) $
      &$i\Gamma^\prime_M$
      \\
      \hline
 pseudoscalar ($^1S_0$)
      &$H^\prime_P\gamma_5$
      \\
 vector ($^3S_1$)
      &$i H^\prime_V [\gamma_\mu-\frac{1}{W^\prime_V}(p^\prime_1-p_2)_\mu]$
      \\
 axial ($^3 P_1$)
      &$-i H^\prime_{^3\!A}[\gamma_\mu+\frac{1}{W^\prime_{^3\!A}}(p^\prime_1-p_2)_\mu]\gamma_5$
      \\
 axial ($^1 P_1$)
      &$-i H^\prime_{^1\!A} [\frac{1}{W^\prime_{^1\!A}}(p^\prime_1-p_2)_\mu]\gamma_5$
      \\
 tensor ($^3P_2$)
      &$i\frac{1}{2} H^\prime_T [\gamma_\mu-\frac{1}{W^\prime_V}(p^\prime_1-p_2)_\mu](p^\prime_1-p_2)_\nu$
      \\
\hline
\end{tabular}
\end{table}

%
We need Feynman rules for the meson-quark-antiquark vertices to
calculate the amplitudes depicted in Fig.~1. The Feynman rules for
vertices ($i\Gamma^\prime_M$) of ground-state $s$-wave mesons and
low-lying $p$-wave mesons are summarized in Table~\ref{tab:feyn}.
Note that we use $^3A$ and $^1A$ to denote $^3P_1$ and $^1P_1$
states, respectively. It is known that the integration of the
minus component of the internal momentum in Fig.~1 will force the
antiquark to be on its mass shell \cite{Jaus99}. The specific form
of the (phenomenological) covariant vertex functions for on-shell
quarks can be determined by comparing to the conventional vertex
functions~\cite{CCH}.

We first consider the tensor form factors for $B_q\to V$
transition. We have
 \be
 {\cal B}_{\mu\nu}\vp^{\pp*\nu}
 \equiv\la V(P^\pp,\vp^\pp)|\bar s\sigma_{\mu\lambda} q^\lambda
 (1+\gamma_5) b|\overline B_q(P^\prime)\ra
 =-i^3\frac{N_c}{(2\pi)^4}\int d^4 p^\prime_1
 \frac{H^\prime_P (i H^\pp_V)}{N_1^\prime N_1^\pp N_2} S_{R\mu\nu}\,\vp^{\pp*\nu},
 \label{eq:B}
 \en
where
 \be
S_{R\mu\nu} ={\rm
Tr}\left[\left(\gamma_\nu-\frac{1}{W^\pp_V}(p_1^\pp-p_2)_\nu\right)
                                 (\not \!p^\pp_1+m_1^\pp)
                                 \sigma_{\mu\lambda} q^\lambda (1+\gamma_5)
                                 (\not \!p^\prime_1+m_1^\prime)\gamma_5(-\not
                                 \!p_2+m_2)\right],
 \en
$N_1^\pp=p_1^{\pp2}-m_1^{\pp2}+i\epsilon$ and
$N_2=p_2^2-m_2^2+i\epsilon$. By using the identity
$2\sigma_{\mu\lambda}\gamma_5=i\epsilon_{\mu\lambda\rho\sigma}\sigma^{\rho\sigma}$,
the above trace $S_{R\mu\nu}$ can be further decomposed into
 \be
S_{R\mu\nu}=q^\lambda
S_{\nu\mu\lambda}+\frac{i}{2}q^\lambda\epsilon_{\mu\lambda\rho\sigma}
S_\nu^{\,\,\rho\sigma}.
 \en
It is straightforward to show that
 \be
 S_{\nu\mu\lambda}&=&
 2\epsilon_{\mu\nu\alpha\lambda}
 \left[2(m^\prime_1 m_2+m_1^\pp m_2-m^\prime_1 m_1^\pp) p^{\prime\alpha}_1
  +m_1^\prime m_1^\pp P^\alpha
  +(m_1^\prime m_1^\pp-2 m_1^\prime m_2) q^\alpha
 \right]
 \non\\
&&-\frac{1}{W^\pp_V}(4 p^\prime_{1\nu}-3
           q_\nu-P_\nu)\epsilon_{\mu\lambda\alpha\beta}
           [(m_1^\prime+m_1^\pp)p^{\prime\alpha}_1 P^\beta
           +(m_1^\pp-m_1^\prime+2m_2) p_1^{\prime\alpha} q^\beta
 +m_1^\prime P^\alpha q^\beta]\non \\
 && +\Big\{2\epsilon_{\mu\nu\alpha\lambda}
 \big[2(p'_1\cdot p_2-p''_1\cdot p_2-p'_1\cdot p''_1) p^{\prime\alpha}_1
  +p'_1\cdot p''_1 P^\alpha
  +(-2p'_1\cdot p_2+p'_1\cdot p''_1) q^\alpha
 \big]
 \non\\
 &&
 +2(g_{\lambda\nu}\epsilon_{\mu\alpha\beta\rho}-g_{\mu\nu}\epsilon_{\lambda\alpha\beta\rho})
 P^\alpha q^\beta p^{\prime\rho}_1
 +2\epsilon_{\lambda\mu\alpha\beta}
 (P^\alpha q^\beta p'_{1\nu}+p^{\prime\alpha}_1 P^\beta q_{\nu}+q^\alpha p^{\prime\beta}_1 P_{\nu})
 \non\\
 &&+
 2\epsilon_{\mu\nu\alpha\beta}
 \big[
 p'_{1\lambda}P^\alpha q^\beta+q_\lambda P^\alpha p^{\prime\beta}_1
 +(P+2q)_\lambda q^\alpha p^{\prime\beta}_1
 +2p'_{1\lambda} p^{\prime\alpha}_1 (P+q)^\beta
 \big]
 \non\\
 &&-2\epsilon_{\lambda\nu\alpha\beta}
 \big[
 p'_{1\mu}P^\alpha q^\beta
 +q_\mu P^\alpha p^{\prime\beta}_1
 +(P+2q)_\mu q^\alpha p^{\prime\beta}_1
 +2p'_{1\mu} p^{\prime\alpha}_1 (P+q)^\beta
 \big] \Big\}.
 \label{eq:S}
 \en
Note that those terms in $\{\cdots\}$ are missed in the
original version of~\cite{Cheng:2004yj}.
To proceed, it is useful to use the following identities
 \be
 2p'_1\cdot p_2&=&M'^2-p'^2_1-p_2^2=M'^2-N'_1-N_2-m'^2_1-m_2^2,
 \non\\
 2p''_1\cdot p_2&=&M''^2-p''^2_1-p_2^2=M''^2-N''_1-N_2-m''^2_1-m_2^2,
 \non\\
 2p'_1\cdot
 p''_1&=&-q^2+p'^2_1-p''^2_1=-q^2+N'_1+N''_1+m'^2_1+m''^2_1.
 \en

As in \cite{Jaus99,CCH}, we shall work in the $q^+=0$ frame. For the
integral in Eq.~(\ref{eq:B}) we perform the $p_1^-$
integration~\cite{Jaus99}, which picks up the residue at $p_2=\hat
p_2$ and leads to
 \be
 N_1^{\prime(\pp)}
      &\to&\hat N_1^{\prime(\pp)}=x_1(M^{\prime(\pp)2}-M_0^{\prime(\pp)2}),
\non\\
 H^{\prime(\pp)}_M
      &\to& h^{\prime(\pp)}_M,
\non\\
 W^\pp_M
      &\to& w^\pp_M,
\non\\
\int \frac{d^4p_1^\prime}{N^\prime_1 N^\pp_1 N_2}H^\prime_P
H^\pp_V S
      &\to& -i \pi \int \frac{d x_2 d^2p^\prime_\bot}
                             {x_2\hat N^\prime_1
                             \hat N^\pp_1} h^\prime_P h^\pp_V \hat S,
 \label{eq:contourB}
 \en
where
 \be
 M^{\pp2}_0
          =\frac{p^{\pp2}_\bot+m_1^{\pp2}}
                {x_1}+\frac{p^{\pp2}_{\bot}+m_2^2}{x_2},
 \en
with $p^\pp_\bot=p^\prime_\bot-x_2\,q_\bot$. The explicit forms of
$h^\prime_M$ and $w^\prime_M$ are given by~\cite{CCH}
\begin{eqnarray} \label{eq:vertex}
 h^\prime_P&=&h^\prime_V
                  =(M^{\prime2}-M_0^{\prime2})\sqrt{\frac{x_1 x_2}{N_c}}
                    \frac{1}{\sqrt{2}\widetilde M^\prime_0}\varphi^\prime,
 \nonumber\\
 h^\prime_{^3\!A}
                  &=&(M^{\prime2}-M_0^{\prime2})\sqrt{\frac{x_1 x_2}{N_c}}
                    \frac{1}{\sqrt{2}\widetilde M^\prime_0}\frac{\widetilde
                     M^{\prime
                     2}_0}{2\sqrt{2}M^\prime_0}\varphi^\prime_p,
                        \nonumber\\
 h^\prime_{^1\!A}&=& h^\prime_T =(M^{2\prime}-M_0^{\prime 2})\sqrt{\frac{x_1
 x_2}{N_c}}\frac{1}{\sqrt{2}\widetilde M^\prime_0}\varphi'_p\, ,
 \non\\
 w^\prime_V&=&M^\prime_0+m^\prime_1+m_2,\quad
 w^\prime_{^3\!A}=\frac{\widetilde{M}'^2_0}{m^\prime_1-m_2},\quad
 w^\prime_{^1\!A}=2\,,
 \label{eq:h}
\end{eqnarray}
where $\varphi'$ and $\varphi'_p$ are the light-front momentum
distribution amplitudes for $s$-wave and $p$-wave mesons,
respectively. The Gaussian-type wave function is used \cite{Gauss}
\begin{eqnarray} \label{eq:Gauss}
 \varphi^\prime
    &=&\varphi^\prime(x_2,p^\prime_\perp)
             =4 \left({\pi\over{\beta^{\prime2}}}\right)^{3\over{4}}
               \sqrt{{dp^\prime_z\over{dx_2}}}~{\rm exp}
               \left(-{p^{\prime2}_z+p^{\prime2}_\bot\over{2 \beta^{\prime2}}}\right),
\nonumber\\
 \varphi^\prime_p
    &=&\varphi^\prime_p(x_2,p^\prime_\perp)=\sqrt{2\over{\beta^{\prime2}}}~\varphi^\prime,\quad\qquad
         \frac{dp^\prime_z}{dx_2}=\frac{e^\prime_1 e_2}{x_1 x_2 M^\prime_0}.
 \label{eq:wavefn}
\end{eqnarray}
The parameter $\beta'$ is expected to be of order $\Lambda_{\rm
QCD}$.

In general, $\hat p^\prime_1$ can be expressed in terms of three
external vectors, $P^\prime$, $q$ and $\tilde\omega$
[$\tilde\omega$ being a lightlike vector with the expression
$\tilde\omega^\mu=(\tilde\omega^-,\tilde\omega^+,\tilde\omega_\bot)=(2,0,0_\bot)$].
In practice, for $\hat p_1^\prime $ under integration we use the
following rules~\cite{Jaus99}
 \be
\hat p^\prime_{1\mu}
       &\doteq& P_\mu A_1^{(1)}+q_\mu A_2^{(1)},
 \non\\
\hat p^\prime_{1\mu} \hat p^\prime_{1\nu}
       &\doteq& g_{\mu\nu} A_1^{(2)}+P_\mu P_\nu A_2^{(2)}+(P_\mu
                q_\nu+ q_\mu P_\nu) A^{(2)}_3+q_\mu q_\nu A^{(2)}_4,
 \non\\
\hat p^\prime_{1\mu} \hat p^\prime_{1\nu} \hat p^\prime_{1\alpha}
       &\doteq& (g_{\mu\nu} P_\alpha+g_{\mu\alpha} P_\nu+g_{\nu\alpha} P_\mu) A_1^{(3)}
               +(g_{\mu\nu} q_\alpha+g_{\mu\alpha} q_\nu+g_{\nu\alpha} q_\mu) A_2^{(3)}
 \non \\
       &&       +P_\mu P_\nu P_\alpha A_3^{(3)}
                +(P_\mu P_\nu q_\alpha+ P_\mu q_\nu P_\alpha+q_\mu P_\nu P_\alpha) A^{(3)}_4
 \non \\
       &&       +(q_\mu q_\nu P_\alpha+ q_\mu P_\nu q_\alpha+P_\mu q_\nu q_\alpha)
                 A^{(3)}_5
                +q_\mu q_\nu q_\alpha  A^{(3)}_6,
 \non\\
 \hat N_2&\doteq& Z_2,
 \non\\
 \hat p_{1\mu}' \hat N_2&\doteq& q_\mu\left[A^{(1)}_2 Z_2+\frac{P\cdot q}{q^2}
 A^{(2)}_1\right],
 \label{eq:p1B}
 \en
where the symbol $\doteq$ reminds us that the above equations are
true only after integration. In the above equation, $A^{(i)}_j$
and $Z_2$ are functions of $x_{1,2}$, $p^{\prime2}_\bot$,
$p^\prime_\bot\cdot q_\bot$ and $q^2$, and their explicit
expressions are given by~\cite{Jaus99}
 \be \label{eq:Aij}
 A^{(1)}_1&=&\frac{x_1}{2},
 \quad
 A^{(1)}_2=A^{(1)}_1-\frac{p^\prime_\bot\cdot q_\bot}{q^2},
\non\\
 A^{(2)}_1&=&-p^{\prime2}_\bot-\frac{(p^\prime_\bot\cdot q_\bot)^2}{q^2},
 \quad
 A^{(2)}_2=\big(A^{(1)}_1\big)^2,
 \quad
 A^{(2)}_3=A^{(1)}_1 A^{(1)}_2,
 \non\\
 A^{(2)}_4&=&\big(A^{(1)}_2\big)^2-\frac{1}{q^2}A^{(2)}_1,
 \quad
 A^{(3)}_1=A^{(1)}_1 A^{(2)}_1,
 \quad
 A^{(3)}_2=A^{(1)}_2 A^{(2)}_1,
 \\
 A^{(3)}_3&=&A^{(1)}_1 A^{(2)}_2,
 \quad
 A^{(3)}_4=A^{(1)}_2 A^{(2)}_2,
 \quad
 A^{(3)}_5=A^{(1)}_1 A^{(2)}_4,
  \non\\
 A^{(3)}_6&=&A^{(1)}_2 A^{(2)}_4-\frac{2}{q^2}A^{(1)}_2 A^{(2)}_1,
  \non\\
 Z_2&=&\hat N'_1+m_1^{\prime 2}-m_2^2+(1-2x) M^{\prime 2}+(q^2+q\cdot P)
 \frac{p'_\bot q_\bot}{q^2}.
 \en

The calculation for $B_q\to A_{^3P_1,^1P_1}$ transition form
factors can be done in a similar manner. In analogue to
Eq.~(\ref{eq:B}), we have
 \be
 {\cal B}^{^3P_1}_{\mu\nu}\vp^{\pp*\nu}
  &=&-i^3\frac{N_c}{(2\pi)^4}\int d^4 p^\prime_1
 \frac{H^\prime_P (-i H^\pp_{^3A})}{N_1^\prime N_1^\pp N_2} S^{^3A}_{R\mu\nu}\,\vp^{\pp*\nu},
 \non\\
 {\cal B}^{^1P_1}_{\mu\nu}\vp^{\pp*\nu}
  &=&-i^3\frac{N_c}{(2\pi)^4}\int d^4 p^\prime_1
 \frac{H^\prime_P (-i H^\pp_{^1A})}{N_1^\prime N_1^\pp N_2} S^{^1A}_{R\mu\nu}\,\vp^{\pp*\nu},
 \label{eq:Baxial}
 \en
where
 \be
S^{^3A}_{R\mu\nu} &=&{\rm
Tr}\left[\left(\gamma_\nu-\frac{1}{W^\pp_{^3A}}(p_1^\pp-p_2)_\nu\right)\gamma_5
                                 (\not \!p^\pp_1+m_1^\pp)
                                 \sigma_{\mu\lambda} q^\lambda (1+\gamma_5)
                                 (\not \!p^\prime_1+m_1^\prime)\gamma_5(-\not
                                 \!p_2+m_2)\right],
 \non\\
S^{^1A}_{R\mu\nu} &=&{\rm
Tr}\left[\left(-\frac{1}{W^\pp_{^1A}}(p_1^\pp-p_2)_\nu\right)\gamma_5
                                 (\not \!p^\pp_1+m_1^\pp)
                                 \sigma_{\mu\lambda} q^\lambda (1+\gamma_5)
                                 (\not \!p^\prime_1+m_1^\prime)\gamma_5(-\not
                                 \!p_2+m_2)\right].
 \en
It can be easily shown that $S^{^3A,^1A}_{R\mu\nu}=-S_{R\mu\nu}$
with $m_1^\pp$ and $W^\pp_V$ replaced by $-m_1^\pp$ and
$W^\pp_{^3A,^1A}$, respectively, while only the $1/W^\pp_{^1A}$
term is kept for the $S^{^1A}_R$ case. Consequently, we have, for
$i=1,2,3$,
 \be\label{eq:TA}
 Y_{Ai,Bi}(q^2)&=&T_i(q^2) \,\,\,{\rm with}\,\,\,
                         (m_1^\pp\to -m_1^\pp,\,h^\pp_V\to h^\pp_{^3\!A,^1\!A},\,w^\pp_V\to w^\pp_{^3\!A,^1\!A}),
 \en
where only the $1/W^\pp$ terms in $Y_{Bi}$ form factors are kept.
It should be cautious that the replacement of $m_1^\pp\to
-m_1^\pp$ should not be applied to $m_1^\pp$ in $w^\pp$ and
$h^\pp$.

Finally we turn to the $B_q\to T$ transition given by
 \be
 {\cal B}^{T}_{\mu\nu\lambda}\vp^{\pp*\nu\lambda}
 \equiv
 \la T(P^\pp,\vp^\pp)|\bar s\sigma_{\mu\nu}(1+\gamma_5) q^\nu b|\bar B_q(P^\prime)\ra
 =-i^3\frac{N_c}{(2\pi)^4}\int d^4 p^\prime_1
 \frac{H^\prime_P (-i H^\pp_T)}{N_1^\prime N_1^\pp N_2} S^{PT}_{\mu\nu\lambda}\,\vp^{\pp*\nu\lambda},
 \label{eq:BT}
 \en
where
 \be
 S^T_{R\mu\nu\lambda} \vp^{\pp*\nu\lambda}(p'')\equiv
 S_{R\mu\nu}\frac{(p_2-p''_1)_\lambda}{2}
 \vp^{\pp*\nu\lambda}(p'')
 =S_{R\mu\nu}(q-p'_1)_\lambda \vp^{\pp*\nu\lambda}(p'').
 \label{eq:ST}
 \en
The contribution from the $S_{\mu\nu}q_\lambda$ part is trivial,
since $q_\lambda$ can be taken out from the integration, which is
already done in the $B_q\to V$ case. Contributions from the $\hat
S_{R\mu\nu}\hat p^\prime_{1\lambda}$ part can be worked out by
using Eq.~(\ref{eq:p1B}).

The final results of these calculations, i.e. tensor form factors
for $B_q\to M$ transitions, are given in~\cite{Cheng:2004yj} and
recollected in Sec. II.

\section{Input parameters for decay amplitudes in the QCDF approach}

\begin{table}[t]
\caption{ Input parameters. The values of the scale dependent
quantities $f^\perp(\mu_h)$ and $a^{\bot}_{0,1,2}(\mu_h)$ are
given for $\mu_h=1\,\rm{GeV}$.
} \label{tab:QCDFinput}
\begin{center}
\begin{tabular}{|c||c|c|c|c|}
\hline\hline
\multicolumn{5}{|c|}{Light mesons} \\
 \hline
 $M$
   & $f^\perp_M({\rm MeV})$
   & $a^\bot_0$
   & $a^\bot_1$
   & $a^\bot_2$
   \\
 \hline
 $K^*$~\cite{BJZ}
   & $185\pm 10$
   & 1
   & $0.04\pm0.03$
   & $0.15\pm0.15$
   \\
 $\phi$~\cite{BJZ}
   & $186\pm 9$
   & 1
   & 0
   & $0.2\pm0.2$
   \\
 $K_{1A}$~\cite{Hatanaka}
   & $250\pm 13$
   & $0.26^{+0.03}_{-0.22}$
   & $-1.08\pm 0.48$
   & $0.02\pm0.20$
   \\
 $K_{1B}$~\cite{Hatanaka}
   & $-190\pm 10$
   & 1
   & $0.30^{+0.00}_{-0.31}$
   & $-0.02\pm0.22$
   \\
 $f_1^{^3P_1}$~\cite{Yang:2007zt}
   & $245\pm13$
   & 0
   & $-1.06\pm0.36$
   & 0
   \\
 $f_8^{^3P_1}$~\cite{Yang:2007zt}
   & $239\pm13$
   & 0
   & $-1.11\pm0.31$
   & 0
   \\
 $h_1^{^1P_1}$~\cite{Yang:2007zt}
   & $-180\pm12$
   & $1$
   & 0
   & $0.18\pm0.22$
   \\
 $h_8^{^1P_1}$~\cite{Yang:2007zt}
   & $-190\pm10$
   & $1$
   & 0
   & $0.14\pm0.22$
   \\
 \hline\hline
  \multicolumn{5}{|c|}{$B$ mesons~\cite{PDG}}
   \\
 \hline
  $B$
   & {$m_B({\rm GeV}$)}
   & $\tau_B({\rm ps})$
   & {$f_B({\rm MeV})$}
   & $\lambda_B({\rm MeV})$
   \\
 \hline
  $B_u$
   & {$5.279$}
   & $1.638$
   & {$200\pm 15$}
   & $350\pm 100$
   \\
 \hline
  $B_s$
   & {$5.366$}
   & $1.472$
   & {$230\pm 15$}
   & $350\pm 100$
   \\
\hline\hline
  \multicolumn{5}{|c|}{Form factors $F^{B\to M}(0)$ (this work) }
   \\
 \hline
  \multicolumn{2}{|c|}{$T_1^{B\to K^*}(0)$}
   & $Y_{A1}^{B\to K_{1A}}(0)$
   & $Y_{B1}^{B\to K_{1B}}(0)$
   & $U_1^{B\to K_2}(0)$
   \\
 \hline
  \multicolumn{2}{|c|}{$0.29\pm0.03$}
   & $0.36\pm0.02$
   & $0.13\pm0.01$
   & $0.28\pm0.03$
   \\
 \hline
   \multicolumn{2}{|c|}{$T_1^{B_s\to \phi}(0)$}
   & $Y_{A1}^{B_s\to f^s_{^3P_1}}(0)$
   & $Y_{B1}^{B_s\to h^s_{^1P_1}}(0)$
   & $U_1^{B\to f^s_{^3P_2}}(0)$
   \\
 \hline
  \multicolumn{2}{|c|}{$0.27\pm0.03$}
   & $0.36\pm0.02$
   & $0.12\pm0.01$
   & $0.28\pm0.03$
   \\
 \hline\hline
  \multicolumn{5}{|c|}{Quark masses~\cite{PDG}}
   \\
 \hline
  \multicolumn{3}{|c|}{$m_b(m_b)/{\rm GeV}$}
   & \multicolumn{2}{|c|}{$m_c/m_b$}
   \\
 \hline
  \multicolumn{3}{|c|}{$4.20^{+0.17}_{-0.07}$}
   & \multicolumn{2}{|c|}{$0.31$}
   \\
 \hline\hline
  \multicolumn{5}{|c|}{CKM matrix elements~\cite{CKMfitter}}
   \\
 \hline
  \multicolumn{3}{|c|}{$|V_{cb}|$}
   & \multicolumn{2}{|c|}{$|V_{cs}|$}
   \\
 \hline
  \multicolumn{3}{|c|} {$0.04117^{+0.00038}_{-0.00117}$}
   & \multicolumn{2}{|c|}{$0.97349^{+0.00018}_{-0.00017}$}
   \\
\hline \hline
\end{tabular}
\end{center}
\end{table}

Input parameters of the radiative $B$ decay amplitudes are
collected in Table~\ref{tab:QCDFinput}. Values of form factors are
calculated in this work. Other hadronic parameters are
from~\cite{PDG,BJZ,Yang:2007zt,Hatanaka}. Note that the signs of
$f^\bot_M$ for $M={}^1P_1$ states are flipped to match our sign
convention. For Gegenbauer moments of physical mesons, we use
 \be
 a_i^{\bot,K_1(1270)}&=&\frac{f^\bot_{K_{1A}}}{f_{K_1(1270)}^\bot} a_i^{\bot,K_{1A}}\sin\theta_K
                       +\frac{f^\bot_{K_{1B}}}{f_{K_1(1270)}^\bot} a_i^{\bot,K_{1A}}\cos\theta_K,
 \non\\
 a_i^{\bot,K_1(1400)}&=&\frac{f^\bot_{K_{1A}}}{f_{K_1(1400)}^\bot} a_i^{\bot,K_{1A}}\cos\theta_K
                       -\frac{f^\bot_{K_{1B}}}{f_{K_1(1400)}^\bot} a_i^{\bot,K_{1A}}\sin\theta_K,
 \non\\
 a_i^{\bot,f^s}&=&\frac{f^\bot_{f_1}}{f_f^\bot} a_i^{\bot,f_1}\frac{\cos\theta}{\sqrt3}
                       -2\frac{f_8^\bot}{f_f^\bot} a_i^{\bot,f_8}\frac{\sin\theta}{\sqrt6},
 \non\\
 a_i^{\bot,f^{\prime s}}&=&-\frac{f_1^{\prime\bot}}{f_{f'}^\bot} a_i^{\bot,f'_1}\frac{\sin\theta}{\sqrt3}
                       -2\frac{f_8^{\prime\bot}}{f_{f'}^\bot} a_i^{\bot,f'_2}\frac{\cos\theta}{\sqrt6},
 \en
with
 \be
 f^\bot_{K_1(1270)}&=&f^\bot_{K_{1A}}\sin\theta_K+f^\bot_{K_{1B}}\cos\theta_K,
 \non\\
 f^\bot_{K_1(1400)}&=&f^\bot_{K_{1A}}\cos\theta_K-f^\bot_{K_{1B}^\bot}\sin\theta_K,
 \non\\
 f^\bot_{f^s}&=&f_{f_1}^\bot\frac{\cos\theta}{\sqrt3}-2f^\bot_{f_8}\frac{\sin\theta}{\sqrt6},
 \non\\
 f^\bot_{f^{\prime s}}&=&-f_{f'_1}^{\bot}\frac{\sin\theta}{\sqrt3}
                       -2f_{f'_8}^{\bot}\frac{\cos\theta}{\sqrt6},
 \en
where $\theta=\alpha-54.7^\circ$ and $f,f'$ are the states specified in
Table~\ref{tab:f'f}. The scale $\mu$ for $a^c_7$ is varied from
$m_b/2$ to $2m_b$.

\end{document}